%% file: declock.tex
\documentclass[10pt,letterpaper,twocolumn]{article}
\usepackage[10pt,nocopyright]{sigmin}

\usepackage{listings}
\usepackage{color}
\usepackage{epsfig,makecell}    
\usepackage{ifpdf}
\ifpdf
  \usepackage{epstopdf}
\fi                                  %%
\usepackage[square,comma,numbers,sort&compress]{natbib}          %%
\usepackage[hyphens]{url}                                         %%
\usepackage[colorlinks]{hyperref}                      %%
\usepackage[usenames,dvipsnames]{xcolor}                          %%
\hypersetup{citecolor=blue,linkcolor=blue}                        %%
\usepackage{amsmath,amsopn,amssymb,amsthm}                        %%
\usepackage{thmtools}
\usepackage[toc,page]{appendix}
\usepackage{subfigure}
\usepackage{endnotes,microtype,xspace,fancyvrb,multirow} %%
\usepackage{booktabs}                                             %%
\usepackage{array,underscore,relsize}                             %%
\usepackage[T1]{fontenc}                                          %%
\usepackage{times}                                                %%
\usepackage{fancyhdr,lastpage}                                    %%
\usepackage{enumitem}                                             %%
\usepackage[export]{adjustbox}                                    %%
\usepackage{breakurl}                                             %%
\usepackage{pifont}                                               %%
\usepackage{tablefootnote}                                        %%
\usepackage[linesnumbered,vlined,ruled,commentsnumbered]{algorithm2e}
\DontPrintSemicolon

\SetCommentSty{mycommfont}

\usepackage{amssymb}
\usepackage[normalem]{ulem}
\usepackage[hang,flushmargin]{footmisc}
\usepackage{textcomp}
\usepackage{graphicx}

\usepackage[compact]{titlesec}
\titleformat*{\section}{\large\bfseries}
\titleformat*{\subsection}{\normalsize\bfseries}
\titleformat*{\subsubsection}{\normalsize\bfseries}
\titlespacing*{\section}{0pt}{*2}{*1}
\titlespacing*{\subsection}{0pt}{*1.5}{*0.8}

\definecolor{mygreen}{rgb}{0,0.6,0}  
\definecolor{mygray}{rgb}{0.5,0.5,0.5}  
\definecolor{mymauve}{rgb}{0.58,0,0.82}

\hypersetup{
  colorlinks,
  linkcolor={red!50!black},
  citecolor={blue!50!black},
  urlcolor={blue!80!black}
}

\input{cmds.tex}

\usepackage{balance}
\usepackage{upgreek}

\begin{document}

%\title{Performant, Scalable, and Practical In-network Lock Management}
%\title{\LARGE \bf Rethinking In-network Lock Management}
%\title{\Large \bf Scalable, Efficient and Fair Lock Management on Disaggregated Memory}

\title{\Large \bf {\sys}: A Case of Decoupled Locking for Disaggregated Memory}

% \author{{Paper \#463}\vspace{3mm}}
% \author{Graduate category}

% \authorinfo{Hanze Zhang\and Ke Cheng\and Rong Chen\and Haibo Chen}
          %  {Institute of Parallel and Distributed Systems, Shanghai Jiao Tong University}
          %  {}

\author{
    Hanze Zhang\;\;\;\;
    Ke Cheng\;\;\;\;
    Rong Chen\;\;\;\;
    Xingda Wei\;\;\;\;
    Haibo Chen\\[5pt]
 \normalsize{{Institute of Parallel and Distributed Systems, Shanghai Jiao Tong University}} \\ [15pt]
 %\normalsize{\emph{Contacts: {rongchen}@sjtu.edu.cn}}
}

\maketitle
\frenchspacing

\baselineskip=12pt

\input{abs.tex}

\input{intro.tex}
\input{bg.tex}

\input{overview.tex}

\input{proto.tex}
\input{design.tex}

\input{impl.tex}
\input{eval.tex}

\input{related.tex}

\input{concl.tex}

% The 'abbrvnat' bibliography style is recommended.

\balance

\small{
  \baselineskip=12pt
  \bibliographystyle{plain}
  \bibliography{bib/dblp,bib/misc}
}

\newpage
\input{appendix.tex}

\clearpage

\end{document}

%% file: cmds.tex
\newcommand{\squishlist}{
  \begin{list}{$\bullet$}{
    \setlength{\itemsep}{0pt}
    \setlength{\parsep}{3pt}
    \setlength{\topsep}{3pt}
    \setlength{\partopsep}{0pt}
    \setlength{\leftmargin}{1.5em}
    \setlength{\labelwidth}{1em}
    \setlength{\labelsep}{0.5em}
  }
}

\newcommand{\squishend}{
  \end{list}
}

\newcounter{enumctr}
\newcommand{\squishenum}{
  \begin{list}{\arabic{enumctr}.}{
    \usecounter{enumctr}
    \setlength{\itemsep}{0pt}
    \setlength{\parsep}{3pt}
    \setlength{\topsep}{3pt}
    \setlength{\partopsep}{0pt}
    \setlength{\leftmargin}{1.5em}
    \setlength{\labelwidth}{1em}
    \setlength{\labelsep}{0.5em}
  }
}

\usepackage{tikz}

\newcommand{\stitle}[1]{\vspace{.8ex}\noindent{\bf #1}}
\newcommand{\nospacestitle}[1]{\noindent{\bf #1}}
\newcommand{\etitle}[1]{\vspace{0.5ex}\noindent{\em\underline{#1}}}

\newcommand{\sys}[0]{\textsc{DecLock}\xspace}

\newcommand{\us}{{{$\upmu$}s}\xspace}
\newcommand{\x}{{$\times$}\xspace}
\newcommand{\p}[1]{{{#1}$^\text{th}$-percentile}\xspace}

\let\oldding\ding%
\renewcommand{\ding}[2][1]{\scalebox{#1}{\oldding{#2}}}%

%% file: abs.tex
\begin{abstract}

This paper reveals that locking can significantly degrade the performance
of applications on disaggregated memory (DM), sometimes by several orders of magnitude, 
due to contention on %
the NICs of memory nodes (MN-NICs).
To address this issue, we present {\sys}, a locking mechanism for DM 
that employs decentralized coordination for \emph{ownership transfer} across 
compute nodes (CNs)
while retaining centralized \emph{state maintenance} on memory nodes (MNs).
{\sys} features cooperative queue-notify locking that queues 
lock waiters on MNs atomically, enabling clients to transfer lock 
ownership via message-based notifications between CNs.
This approach conserves MN-NIC resources for DM applications 
and ensures fairness. 
Evaluations show {\sys} achieves throughput improvements of up to 
43.37\x and 1.81\x
over state-of-the-art RDMA-based spinlocks and MCS locks, respectively.
Furthermore, {\sys} helps two DM applications, including an object store
and a real-world database index (Sherman), avoid performance 
degradation under high contention, improving throughput by 
up to 35.60\x and 2.31\x and reducing {\p{99}} latency by 
up to 98.8\% and 82.1\%.

\end{abstract}

%% file: intro.tex
\section{Introduction}

The disaggregated memory (DM) architecture, which detaches CPU and memory
resources into physically disaggregated compute nodes (CNs) and memory 
nodes (MNs), %
attracts increasing interest from both
academia~\cite{legoos,mind,kona} and industry~\cite{dmos,polardbsrvless,ditto}.
Databases~\cite{ford,sherman,smart,polardbsrvless,polardbmp}  
and key-value stores~\cite{clover,dinomo,fusee,ditto,rolex,race}
have been redesigned for DM to allow independently
scaling CPU and memory resources for distributed applications.
In practice, application data on a single MN could be accessed concurrently
by hundreds of clients on multiple CNs~\cite{smart,ditto,chime}, which 
renders the NICs on MNs
(referred to as \textbf{MN-NICs})
a potential performance bottleneck for DM applications.

Numerous DM applications utilize locks stored on MNs to serialize 
conflicting data accesses from different 
CNs~\cite{clover,ford,sherman,smart,rolex,dlsm,marlin,chime}.
To acquire and release these locks, application clients on CNs directly 
manipulate lock states on MNs with remote operations (e.g., one-sided RDMA).
Since the locks are colocated with the data they protect, lock operations 
compete with application data accesses on MN-NICs, amplifying the 
performance impact of locks on DM applications.
Under high-contention workloads---common due to the large number
of concurrent clients~\cite{smart,ditto,chime}---lock operations can easily 
saturate the IOPS resources of MN-NICs, drastically damaging 
application performance.

\begin{figure}[t]
  \begin{minipage}{1.\linewidth}
    \centering\includegraphics[scale=.33]{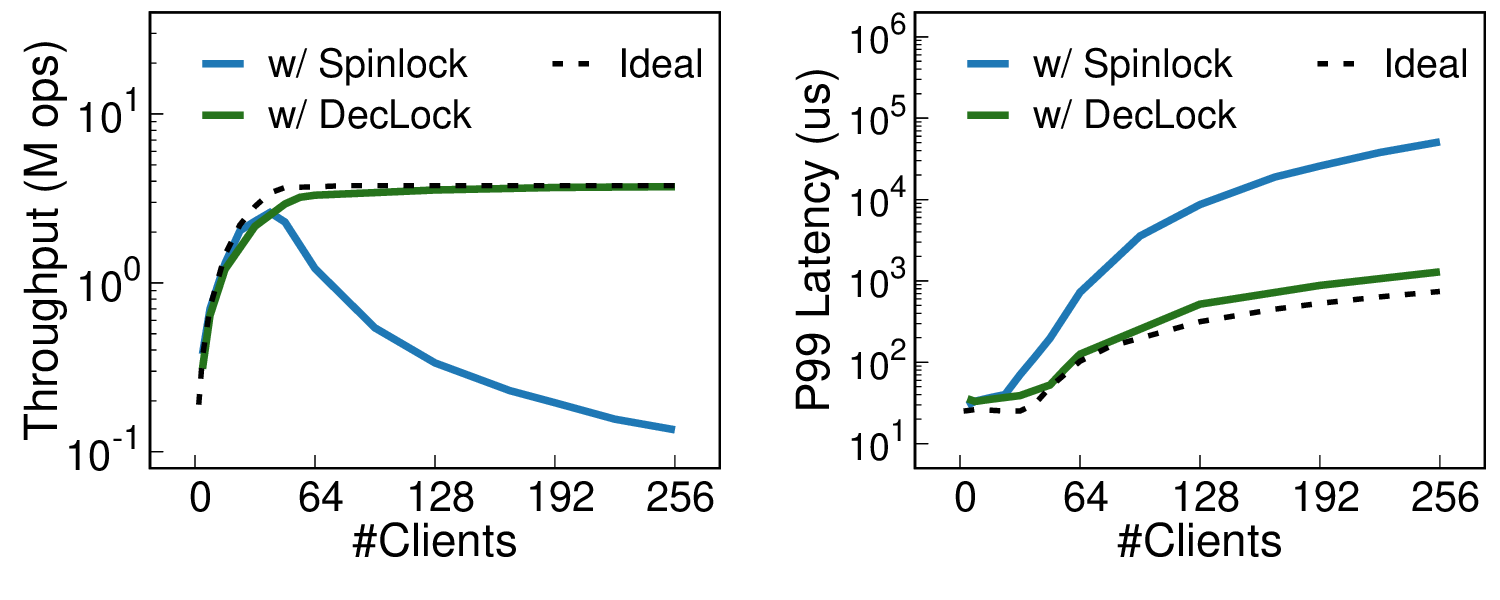}
  \end{minipage} \\[-5pt]
  \begin{minipage}{1.\linewidth}
    \caption{\small{\emph{Throughput and tail latency
	  of update operations in a DM database index 
      using different lock mechanisms
      as \#clients increases.
      \textup{\textbf{Workload:}} 10 million objects w/ a Zipf 
        access distribution. 
      \textup{\textbf{Testbed:}} 8 CNs and 1 MN, connected with 100\,Gbps RDMA NICs.}}}
    \label{fig:intro}
  \end{minipage}  \\[-10pt]
\end{figure}

To demonstrate the performance degradation caused by inefficient locking,
we study the performance of update operations in a database index on DM.
We simulate the ideal performance by running all clients on a single machine 
as coroutines, serializing operations using local locks that incur negligible 
overhead compared to RDMA locks. As shown in Fig.~\ref{fig:intro}, 
when using RDMA spinlocks, the operation throughput collapses to below 5\% 
of the ideal as the number of clients scales to 256, 
and the tail latency surges to 69\x of the ideal.

This significant overhead stems from applying the fail-and-retry 
locking approach---specifically, spinlocks---in DM 
applications~\cite{sherman,smart,ford,rolex}.
To acquire a spinlock, clients must repeatedly retry remote 
operations to check or update lock states on MNs until 
they obtain lock ownership, leading to \textbf{inefficient}
use of MN-NIC IOPS. Worse still, as the number of clients 
increases, retries become more frequent due to longer wait times,
making applications \textbf{unscalable}. For example, in the index
evaluated in Fig.~\ref{fig:intro}, with 256 clients, 
each client averages 28 retries to acquire a lock, consuming 
over 95\% of MN-NIC IOPS.
Moreover, spinlocks are notoriously \textbf{unfair}, which
greatly exacerbates the number of retries for a few unlucky clients,
leading to poor tail latency.

\stitle{{Key idea.}}
We argue that \emph{locking should minimize MN-NIC usage 
to prevent performance interference with applications in disaggregated memory environments.}
Following this principle, we decouple the costly ownership transfer procedure 
from lock state maintenance. This approach enables \textbf{efficient} MN-NIC usage, 
requiring at most two remote operations to acquire a lock.
Ownership transfer leverages decentralized coordination across CNs to enhance \textbf{scalability} 
as the number of clients increases.
Meanwhile, lock states like the identity of waiters remain centralized on MNs to ensure \textbf{fairness}.

\stitle{Our approach.}
We introduce {\sys}, an efficient, scalable, and fair 
reader-writer lock %
that serializes conflicting operations in DM applications 
with near-ideal performance (see Fig.~\ref{fig:intro}).
The core design is the \emph{cooperative queue-notify locking} (CQL) protocol,
which enables lock clients to \emph{enqueue} themselves on MNs and 
\emph{notify} each other between CNs. 
Specifically, when acquiring a lock, the client enqueues itself 
to a centralized array (on MNs) %
and waits for notification from the client that releases the lock (between CNs).
When releasing the lock, the client dequeues itself, retrieves 
information about the next waiter (from MNs), %
and then notifies this waiter to transfer lock ownership (between CNs).

While a centralized queue recording waiters is essential for maintaining
fairness between readers and writers, it is hard to manipulate both the 
lock mode and the queue in a single atomic operation.
To address this challenge, the CQL protocol splits the queue into 
an \emph{atomic control plane} and a \emph{non-atomic data plane}.
The control plane encodes the queue metadata and the lock mode into an 
8-byte atomic header, allowing clients to (de)allocate a queue entry 
and update the lock mode using a single fetch-and-add (FAA) operation. 
Since FAA operations always succeed, this design ensures that readers
and writers obtain the lock in a FIFO manner.
Although the data plane is non-atomic, the atomic allocation of entries
ensures that clients always write to different 
entries in the data plane, eliminating write-write conflicts.
Additionally, the protocol incorporates versions within queue entries 
to identify invalid entries, resolving read-write conflicts.

Although CQL efficiently uses MN-NIC IOPS and preserves fairness,
its centralized queue design can create significant memory and bandwidth 
overhead on MNs for locks acquired by many clients.
Hierarchical locking reduces queue size by enqueuing waiters
at the CN level rather than for individual clients. However, existing
hierarchical designs compromise fairness between waiters on different CNs.
To address this, we employ \emph{timestamp-based hierarchical locking},
which records the acquisition time of each waiting CN in its queue entry. 
When a lock is released, ownership is transferred locally
only if local waiters have an earlier acquisition time than remote CNs,
thereby preserving fairness across CNs.

We evaluated {\sys} using a microbenchmark and two DM applications: 
an object store running real-world traces from Twitter~\cite{twittertrace} 
and a real-world database index (Sherman~\cite{sherman}).
The results show that {\sys} achieves up to 43.37\x and 
1.81\x higher throughput and reduces up to 98.2\% and 43.7\% of tail 
(\p{99}) latency, compared to state-of-the-art RDMA-based 
spinlock~\cite{rdmarw} and MCS lock~\cite{shiftlock}, respectively.
Furthermore, {\sys} effectively addresses throughput collapse of both DM
applications by using MN-NICs efficiently. 
Under high contention, it improves throughput for the 
object store and Sherman by up to 35.60\x and 2.31\x, respectively,
while reducing their tail latency by up to 98.8\% and 82.1\%.

\stitle{Contributions.} We summarize our contributions as follows:

\squishlist
\item A detailed analysis of the performance implications of RDMA-based
  locks on disaggregated memory (\S\ref{sec:bg}).
\item A new \emph{cooperative queue-notify locking} protocol that enables 
  efficient, scalable, and fair lock management on DM by minimizing the usage 
  of MN-NICs (\S\ref{sec:proto}).
\item A new \emph{timestamp-based hierarchical locking} 
  design that allows direct ownership transfer between local clients
  while ensuring fairness across CNs (\S\ref{sec:design}).
\item An evaluation that demonstrates the performance advantages that {\sys}
  offers to DM applications (\S\ref{sec:eval}).
\squishend

%% file: bg.tex
\section{Background and Motivation}\label{sec:bg}

\subsection{DM Architecture and Applications}

\begin{figure}[t]
  \vspace{1mm}
  \begin{minipage}{1.\linewidth}
    \centering\includegraphics[scale=.52]{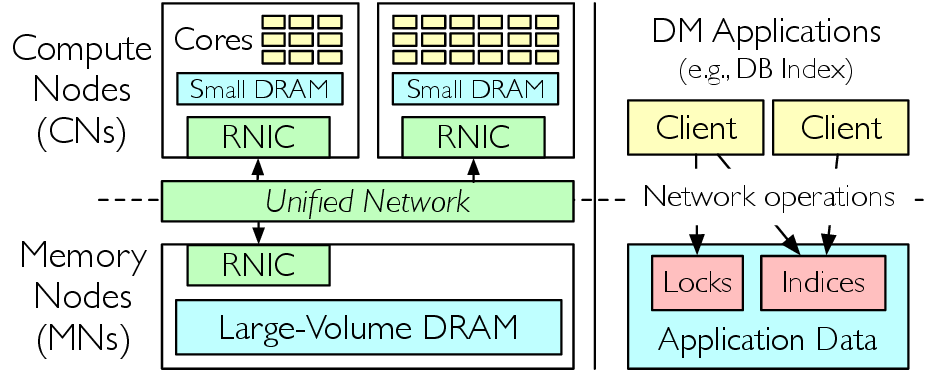}
  \end{minipage} \\[2pt]
  \begin{minipage}{1.\linewidth}
    \caption{\small{\emph{The architecture of DM (left)
    and DM applications (right).
    }}}
    \label{fig:dm}
  \end{minipage}  \\[-15pt]
\end{figure}

The disaggregated memory (DM) architecture is becoming increasingly popular 
in modern data centers~\cite{ford,smart,dmos,ditto}. 
DM separates CPUs and memory into compute nodes (CNs)
and memory nodes (MNs) for better scalability and higher resource utilization 
(see Fig.~\ref{fig:dm} (left)).
CNs possess powerful CPUs for compute tasks and a small piece of DRAM 
(typically a few GB~\cite{fusee,smart}) as cache,
while MNs have large-volume memory for data storage and 
near-zero computing power.
CNs and MNs are typically equipped with RDMA-capable NICs (RNICs) that connect
them with a unified high-speed network that allows direct
communication between any pair of nodes~\cite{firebox,dmnetwork,dmos,legoos}.

Numerous distributed applications have been redesigned on DM, including
transaction processing~\cite{ford,rtx,motor}, 
key-value stores~\cite{clover,dinomo,fusee,ditto,race}, 
and database indices~\cite{sherman,smart,rolex,dlsm,marlin}.
Due to the lack of computing power on MNs, clients heavily rely on 
CPU-bypassing techniques (e.g., one-sided RDMA) to directly access
application data in the MN memory (see Fig.~\ref{fig:dm} (right)).
In practice, RNICs on hotspot MNs often bear significant 
loads generated by hundreds of clients on multiple CNs,
rendering their processing power (IOPS)
a bottleneck for DM applications~\cite{smart,ditto,chime,smart24}. 

To serialize conflicting data accesses, many DM applications rely on locks 
to protect shared data~\cite{ford,clover,sherman,smart,rolex,chime,polardbmp}.
These locks are typically stored on MNs alongside the application data they
protect, allowing for piggybacking of lock operations and data accesses. 
Application clients acquire and release these locks through 
atomic remote operations, such as RDMA compare-and-swap (CAS) and 
fetch-and-add (FAA), 
before and after accessing the protected data.
Since lock operations are on the data access critical path and contend
with data accesses on MN-NICs,
they can drastically degrade the performance of DM applications.
As illustrated in Fig.~\ref{fig:intro}, the CAS-based spinlock adopted
by existing DM applications results in poor scalability and elevated
tail latency under large numbers of clients, which are common in
practice~\cite{smart,ditto,chime,smart24,dinomo}.

\subsection{Performance Issues of DM Spinlock}
\label{sec:bg:spinlock}

Existing DM applications commonly adopt the \emph{spinlock} design 
for its simplicity~\cite{ford,clover,sherman,smart,rolex,chime,motor}.
Spinlock states are 64-bit values that are either 0, indicating 
that the lock is \emph{free}, or a non-zero ID representing the client 
holding the lock (i.e., the \emph{lock holder}).
Clients acquire the lock by swapping the value with their ID using 
CAS operations, which only succeed when the original value is 0.
Therefore, each lock is held by at most one client at a time, with 
other clients (i.e., \emph{lock waiters}) blindly retrying the CAS 
operation until they successfully obtain lock ownership. 
The holder releases the lock by setting the lock value
to 0, which transfers lock ownership to the first client that
successfully swaps the value.
This fail-and-retry mechanism severely impedes application 
performance under high contention workloads, which we demonstrate with
a group of experiments on a microbenchmark.

\stitle{Issue \#1: MN-NIC contention.}
When a spinlock is held, waiters must repeatedly retry the CAS
operation on the lock state to obtain lock ownership.
As more clients compete for locks, the average number of CAS retries
per acquisition rises to up to 37.8 due to longer
wait times (see Fig.~\ref{fig:motiv} (left)).
These retries contend with other RDMA operations for the limited
MN-NIC IOPS resources, leading to elevated median latency for both
lock acquisitions and remote data accesses (see Fig.~\ref{fig:motiv} (middle)).
Consequently, acquisition throughput drops sharply from
a peak of 1.61\,M ops to only 0.20\,M ops.

\stitle{Issue \#2: Poor fairness.}
Fairness among lock clients denotes whether the clients can obtain 
lock ownership in the same (or a similar) order as when they initiated the
acquisition. CAS-based spinlocks cannot enforce fairness, as their ownership 
can be obtained by any client attempting to acquire the lock when it 
is released.
The lack of fairness leads to a (\p{99}) tail latency that is over 1000\x
higher than the median latency (see Fig.~\ref{fig:motiv} (middle)),
which explains the substantial tail latency of the database index.

\subsection{Existing Scalable and Fair Locks}
\label{sec:bg:prior}

Prior work apply classic lock algorithms from multicore systems to
enhance the scalability and fairness of RDMA 
locks~\cite{ddlm,dslr,shiftlock,smart24,alock}. However, none
of these approaches comprehensively address the above 
issues of locking on DM.

\stitle{Ticket locks.}
Ticket locks track the order of clients using a ticket value that 
is atomically incremented with each acquisition~\cite{bakery,dslr}.
By maintaining separate tickets for readers and writers, ticket locks
improve scalability by allowing multiple readers to hold the lock 
simultaneously, and achieve strongest fairness where readers and 
writers obtain ownership in exactly the same order as their 
acquisitions (i.e., \emph{task-fair}~\cite{phasefair}).
Nevertheless, waiters still need to repeatedly read the latest ticket 
value from MNs, leading to contention on MN-NICs. 
While this contention can be alleviated by adding 
delays between reads (i.e., \emph{backoff}~\cite{backoff,smart24}), 
tuning the backoff interval to suit all levels of 
contention is notoriously hard~\cite{nucalock}. 

\begin{figure}[t]
  \begin{minipage}{1.\linewidth}
    \centering\includegraphics[scale=.33]{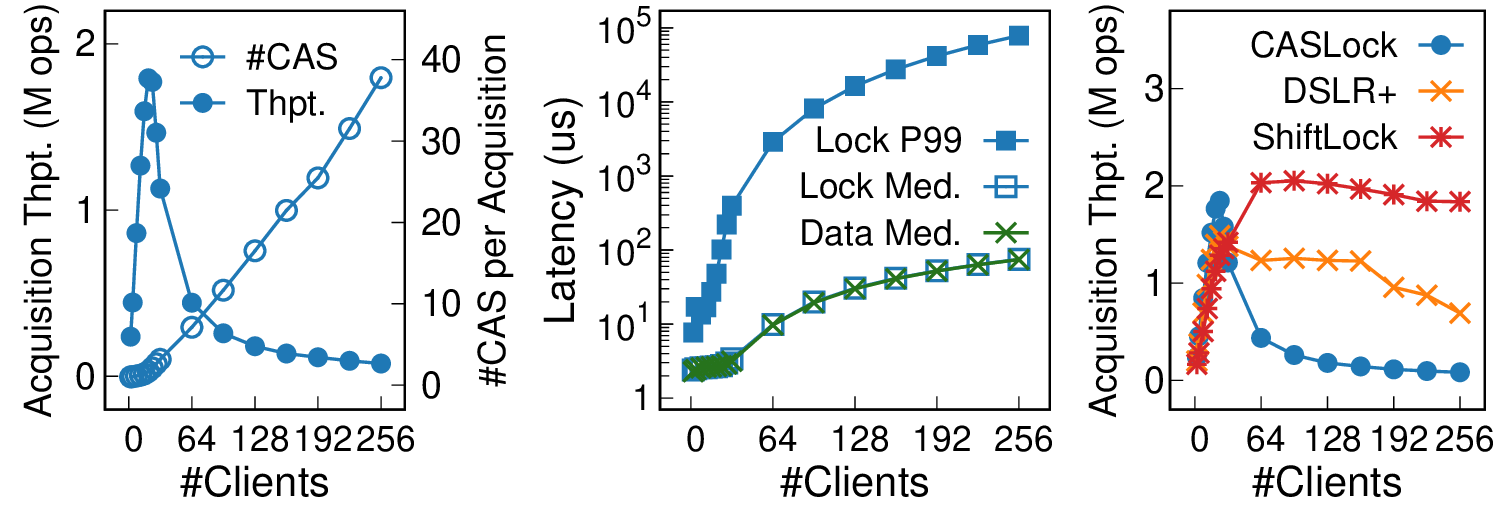}
  \end{minipage} \\[-5pt]
  \begin{minipage}{1.\linewidth}
    \caption{\small{\emph{%
    Acquisition throughput of spinlock and the average number
    of CAS operations per acquisition (left),
    Median and \p{99} latency of spinlock acquisition 
	  and median data access latency (middle), and 
    Acquisition throughput of different lock mechanisms (right).
	Detailed experimental setup can be found in \S\ref{sec:eval}.
    }}}
    \label{fig:motiv}
  \end{minipage}  \\[-10pt]
\end{figure}

To demonstrate this, we integrated the latest backoff technique 
on DM---truncated exponential backoff~\cite{smart24}---into the
state-of-the-art RDMA ticket lock called DSLR~\cite{dslr}.
We compared the resulting lock (DSLR+) with CAS-based spinlocks 
(CASLock) using a workload with an even mix of readers and writers. 
As shown in Fig.~\ref{fig:motiv} (right), DSLR+ maintains 
stable throughput with up to 160 clients because it retries
less frequently (up to 27.4 per acquisition), and read
operations consume less MN-NIC IOPS than CAS operations.
However, it still saturates MN-NIC IOPS and suffers throughput 
degradation when more clients are added. Additionally, the use of
backoff limits the peak throughput of DSLR+.

\stitle{MCS locks.}
MCS locks~\cite{mcs,alock,shiftlock} record waiters in a linked list, 
with the tail pointer 
embedded in the lock state. A recent concurrent work, 
ShiftLock~\cite{shiftlock},
also identifies NIC contention issues in RDMA locks and adopts MCS
locks to reduce NIC usage. After joining the list, waiters can request lock
ownership directly from their predecessors instead of repeatedly checking
lock states on MNs, significantly reducing MN-NIC usage and
alleviating throughput collapse compared to CASLock and DSLR+, 
as shown in Fig.~\ref{fig:motiv} (right).

However, since linked waiters are scattered across CNs and are invisible 
to each other, readers cannot join the list without blocking other 
readers~\cite{shiftlock}. As a result, ShiftLock tracks readers 
using a counter in the lock state, which still requires repeated checks 
when waiting for ownership. To reduce these checks, ShiftLock transfers 
lock ownership to all waiting readers after several consecutive transfers 
between writers. While this approach reduces reader wait times,
it results in weaker fairness (\emph{phase-fair}~\cite{phasefair}) 
and can cause significant latency variance in mixed read-write 
workloads~\cite{phasefair2}.
Despite these efforts, ShiftLock still averages 2.34 checks per acquisition, 
which limits throughput scaling when MN-NIC IOPS is saturated
(see Fig.~\ref{fig:motiv} (right)).
Moreover, under workloads with longer critical sections---common in
scenarios like transaction processing~\cite{ford,tpcc}---it may require
dozens of checks, further increasing MN-NIC contention.

%% file: overview.tex
\section{Approach and Overview}\label{sec:ov}

\begin{figure}[t]
  \begin{minipage}{1.\linewidth}
    \centering\includegraphics[scale=.51]{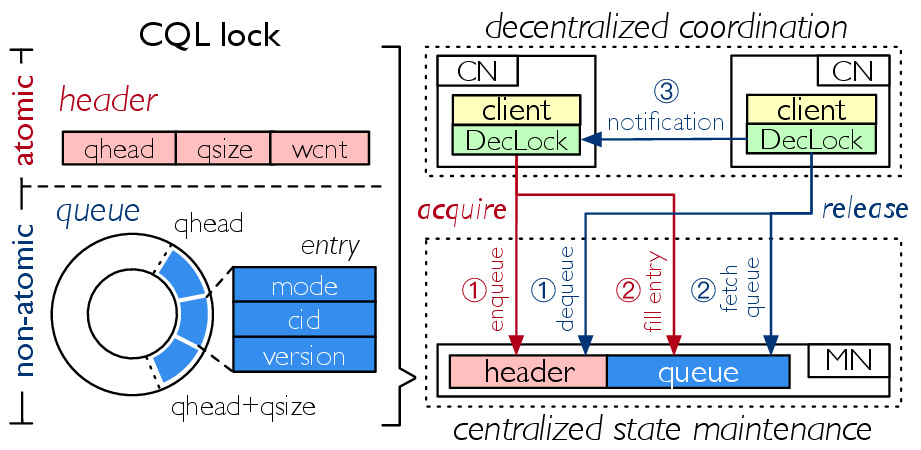}
  \end{minipage} \\[-1pt]
  \begin{minipage}{1.\linewidth}
    \caption{\small{\emph{Architecture, data structure, and workflow of {\sys}.
    }}}
    \label{fig:qlock}
  \end{minipage}  \\[-20pt]
\end{figure}

MN-NICs, responsible for executing remote operations from numerous clients,
have emerged as a significant performance bottleneck in DM applications.
Hence, we propose a fundamental design principle for the lock service on DM:
\emph{minimize the use of MN-NICs} to reduce contention with applications.

\stitle{Our approach.}
Following the principle, we propose \emph{cooperative queue-notify locking} 
(CQL), which decouples ownership transfer from state maintenance to
thoroughly eliminate repeated operations toward MNs while preserving
strong fairness between waiting readers and writers.
Lock states are maintained as a \emph{centralized queue} on MNs that orders
all waiting clients, enabling ownership transfer through 
\emph{decentralized coordination}---notifying waiters  
using direct messages between CNs. Thereby, MNs are accessed only for 
updating lock mode and enqueuing or dequeuing clients.

The CQL design necessitates maintaining multiple lock states, including 
lock mode (free/shared/exclusive), queue metadata, and queue entries, 
which cannot be updated with a single remote atomic operation. 
To tackle this challenge, CQL locks are split into an atomic control 
plane (an 8-byte \emph{header}) and a non-atomic data plane 
(a circular \emph{queue} of clients), as shown in Fig.~\ref{fig:qlock} (left).
The control plane records the used part of the queue 
(\emph{qhead} and \emph{qsize}) and the number of writers in the queue 
(\emph{wcnt}), which implicitly represent the lock mode.
The data plane consists of several queue entries, each storing the acquisition 
mode (\emph{mode}), the client ID (\emph{cid}), and a \emph{version}
for resolving races on the entry 
(see details in \S\ref{sec:proto:version}).

As shown in Fig.~\ref{fig:qlock} (right),
clients acquire the lock using an FAA operation on the header, which
atomically enqueues themselves and returns the original header value
to determine the lock mode ({\ding[1.2]{192}}). If the 
client needs to wait for lock ownership, it subsequently
fills its information into its queue entry ({\ding[1.2]{193}}).
Upon releasing the lock, in addition to dequeuing itself ({\ding[1.2]{192}}),
the client fetches the queue ({\ding[1.2]{193}})
and notifies waiting clients in the queue to
transfer lock ownership ({\ding[1.2]{194}}).

CQL achieves \emph{efficient} usage of MN-NICs by moving the costly lock 
ownership transfer process off the CN-MN interconnect. 
Consequently, lock acquisition requires at most two (and typically just one) 
remote operations towards MNs, leading to substantial savings in MN-NIC 
IOPS compared to existing RDMA locks.
CQL is also highly \emph{scalable} due to its decentralized coordination, 
requiring only one message between CNs for each ownership transfer.
Moreover, CQL is \emph{fair} because both readers and writers are enqueued 
and notified strictly in the order of acquisition.

\stitle{System architecture.}
{\sys} is a lock mechanism that defines the CQL lock 
abstraction and lock operation APIs. %
DM Applications store CQL locks in MN memory and associate them with
shared data they protect, e.g., by embedding them into data entries.
Application clients on CNs call lock operation APIs of {\sys} for 
acquiring and releasing the locks.
{\sys} updates the CQL lock through
RDMA operations towards MNs and transfers the lock ownership through
RDMA messages across CNs.
Notably, while Fig.~\ref{fig:qlock} contains one MN,
{\sys} can seamlessly scale to multiple MNs by distributing CQL locks
to different MNs.

%% file: proto.tex
\section{CQL Protocol}
\label{sec:proto}

\subsection{Lock Header Encoding}\label{sec:proto:encoding}

The CQL lock header is a 64-bit value that is always updated atomically
with FAA operations. Therefore, overflows of particular fields cannot be
detected or prevented, and the only allowed operation on the header is
addition. The lock header encoding is designed to overcome these limitations.

\begin{figure}[t]
  \begin{minipage}{1.\linewidth}
    \centering\includegraphics[scale=.45]{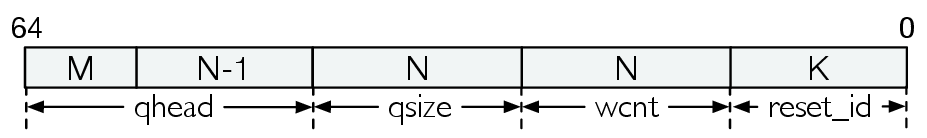}
  \end{minipage} \\[-2pt]
  \begin{minipage}{1.\linewidth}
    \caption{\small{\emph{The CQL lock header encoding.}}}
    \label{fig:qlockhdr}
  \end{minipage}  \\[-15pt]
\end{figure}

\stitle{Field choice and order.}
Overflows of fields in less significant bits corrupt the content of 
fields in more significant bits. Therefore, only the most significant 
field in the header is allowed to overflow.
{\sys} describes the occupied part of the circular queue using its head and 
size rather than head and tail because size has an upper limit 
(i.e., the queue capacity), while both head and tail increment infinitely. 
In the lock header encoding (Fig.~\ref{fig:qlockhdr}), \texttt{qhead}
is placed in the most significant bits so that its overflows have no 
influence on other fields. The remaining bits are assigned to \texttt{qsize}
and \texttt{wcnt} as both fields have a limited maximum value.
The least significant bits are reserved as a reset ID, which indicates 
whether the lock is being reset and identifies the CN responsible
for the reset process (see details in \S\ref{sec:proto:reset}).

\stitle{Field size.}
The number of bits assigned to \texttt{qsize} (marked as \texttt{N} in
Fig.~\ref{fig:qlockhdr}) is larger than that required to represent the 
queue capacity by one. This extra bit ensures that the \texttt{qsize} 
field does not overflow even if the queue size temporarily exceeds the 
capacity (i.e., lock queue overflow), which will be fixed later by 
resetting the lock (detailed in \S\ref{sec:proto:reset}).
The same number of bits are assigned to \texttt{wcnt} because the number 
of exclusive entries in the queue never exceeds the queue size. 
The most significant bits are assigned to \texttt{qhead}, whose value keeps
increasing and is only wrapped within the queue capacity when being used
as the queue index. In addition to the \texttt{N-1} bits within the
queue capacity, \texttt{qhead} has \texttt{M} extra bits indicating
the number of times that the circular buffer has been traversed, which 
serves as the version of queue entries when they are populated
(\S\ref{sec:proto:version}). The number of bits assigned to
the reset ID (\texttt{K}) is sufficient to identify all CNs.

\subsection{Lock Operations}\label{sec:proto:workflow}

\begin{figure}[t]
  \begin{minipage}{1.\linewidth}
    \centering\includegraphics[scale=.49]{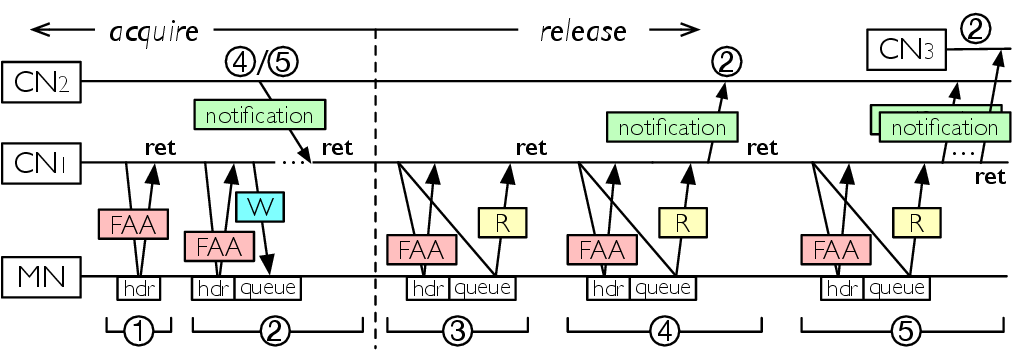}
  \end{minipage} \\[-2pt]
  \begin{minipage}{1.\linewidth}
    \caption{\small{\emph{Workflows of acquiring and releasing
    a CQL lock. ``FAA'', ``W'', and ``R'' denote RDMA FAA, WRITE,
    and READ operations.
    }}}
    \label{fig:workflow}
  \end{minipage}  \\[-10pt]
\end{figure}

\begin{figure}[t]
  \begin{minipage}{1.\linewidth}
    \centering\includegraphics[scale=.75]{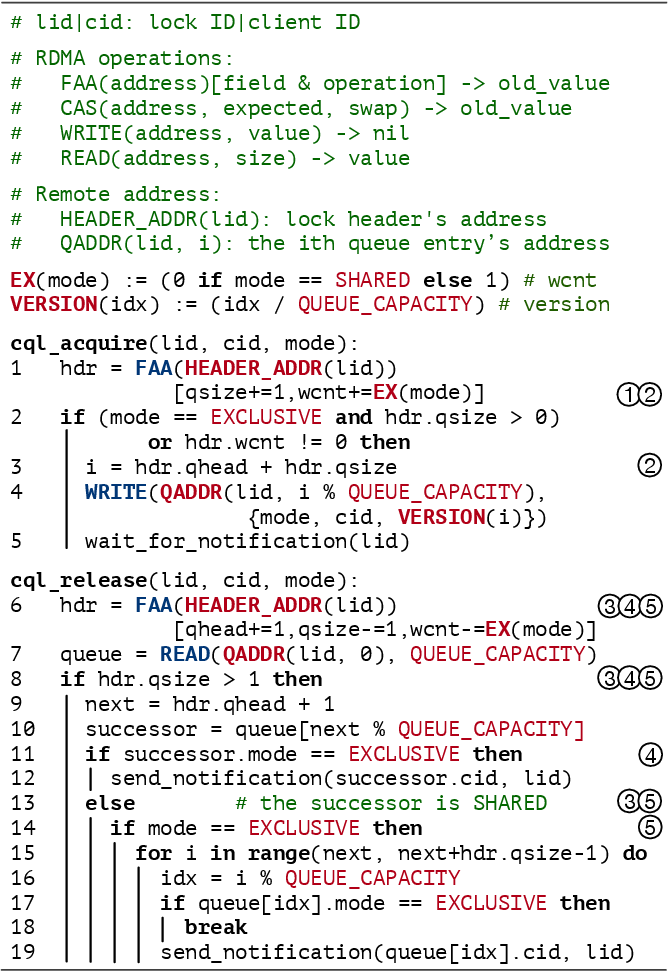}
  \end{minipage} \\[-2pt]
  \begin{minipage}{1.\linewidth}
    \caption{\small{\emph{Pseudocode of CQL acquisition and release
    operations in {\sys}. The destination of all RDMA operations is the
    MN containing the CQL lock, which is omitted for convenience.
    }}}
    \label{fig:qlockop}
  \end{minipage}  \\[-10pt]
\end{figure}

Fig.~\ref{fig:workflow} illustrates the number and type of RDMA
operations used in the five possible workflows of acquiring
({\ding[1.2]{192}}--{\ding[1.2]{193}}) and releasing 
({\ding[1.2]{194}}--{\ding[1.2]{196}}) a CQL lock.
The branches executed in each workflow are marked 
on the pseudocode in Fig.~\ref{fig:qlockop}.
In this context, it is assumed that all queue entries have up-to-date 
content when being read. The mechanism for detecting and rectifying 
obsolete entry contents is introduced in \S\ref{sec:proto:version}.

\stitle{Lock acquisition workflow.}
To acquire a lock, the client first enqueues itself with 
an FAA operation on the lock header, which atomically increments 
\texttt{qsize} by one (Line 1). When acquiring the lock exclusively 
(i.e., when \texttt{EX(mode)} equals 1), the FAA operation 
also increments \texttt{wcnt} by one.
The client then checks the original lock header value returned by 
the FAA operation to determine the result. 
If the lock is acquired in exclusive mode and there are other
clients in the queue, or there are writers in the 
queue, the client becomes a \emph{waiter} of the lock ({\ding[1.2]{193}}).
In this case, the client uses a WRITE operation to populate its 
queue entry with its ID, the acquisition mode, 
and the updated queue entry version (Lines 2--4).
This enables its preceding client to send a notification,
which the current client awaits (Line 5), upon releasing the lock.
If the lock ownership is obtained by the client during this
acquisition process ({\ding[1.2]{192}}), i.e., the client becomes
a \emph{holder} of the lock, the queue entry does not need 
to be populated, as holders do not require notification.

\stitle{Lock release workflow.}
To release an acquired lock, the client dequeues itself using an
FAA operation on the lock header, which simultaneously advances
\texttt{qhead} by one and decrements \texttt{qsize} by one (Line 6).
If the client is a writer, \texttt{wcnt} is also 
decremented by one. {\sys} piggybacks a READ operation, which fetches
the complete lock queue, with this FAA operation to hide the network 
latency (Line 7). If the queue becomes empty after the release
(i.e., the \texttt{qsize} returned by FAA equals 1), the release
operation finishes
as there are no waiters to notify ({\ding[1.2]{194}}). Otherwise, 
{\sys} checks the succeeding entry of the current client's entry
(\texttt{successor}). If the successor is a writer ({\ding[1.2]{195}}), 
its client is certainly waiting for the lock, 
so a notification is sent to wake it up (Lines 11--12). 
If the successor is a reader and the current client is a writer 
({\ding[1.2]{196}}), the lock can be shared by all adjacent readers
in the queue, so all of them are notified (Lines 14--19). 
If both the successor and the current client are readers 
({\ding[1.2]{194}}), the release operation finishes 
without any notification since the succeeding client 
still holds the lock.

\subsection{Queue Entry Versions}\label{sec:proto:version}

When fetching the lock queue during release operations, clients may
encounter obsolete queue entry contents that are populated during 
previous traversals of the lock queue. 
This problem arises from the circular behavior of the lock queue,
which allows previously dequeued entries to be reused by subsequent 
clients. 
To distinguish obsolete entries from valid ones, {\sys} uses the
version of queue entries to represent the freshness of their contents.
The version is determined by the division of the queue index and 
the queue capacity, which represents the number of 
traversals within the circular buffer. Clients update the version 
field when populating their queue entries. When another client reads 
the entry, it compares the version in the entry with the version 
calculated using the queue index it is using. The entry value is 
considered valid only if the two versions are equal.
All entries are initialized with a version value of \texttt{-1},
i.e., all bits in the field are 1.

In the CQL protocol, a fetched entry can be obsolete 
due to read-write conflicts and shared holders.
The protocol handles these situations by \emph{refetching} the obsolete
entries from MNs.
Notably, since both scenarios are rare, the refetching incurs
minimal additional READ operations---at most 1.8\% in our experiments
(see \S\ref{sec:eval:refetch}).

\begin{figure}[t]
  \begin{minipage}{1.\linewidth}
    \centering\includegraphics[scale=.55]{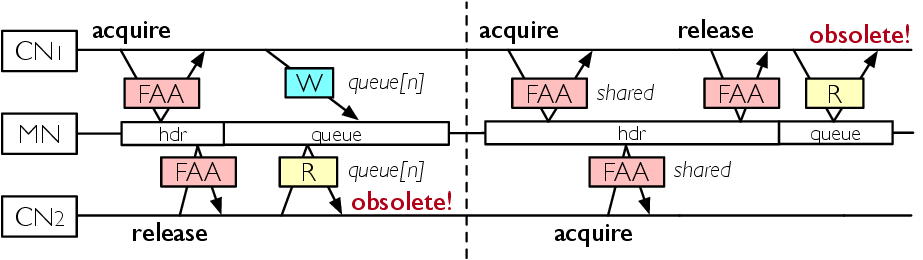}
  \end{minipage} \\[-2pt]
  \begin{minipage}{1.\linewidth}
    \caption{\small{\emph{Two causes of obsolete entries:
      read-write conflicts (left) and shared holders (right).
    }}}
    \label{fig:version}
  \end{minipage}  \\[-15pt]
\end{figure}

\stitle{Read-write conflicts.}
Although the atomic enqueuing mechanism prevents 
write-write conflicts on queue entries, read-write conflicts can occur
due to concurrent acquisition and release operations.
Specifically, when fetching the lock queue during a release, some entries 
might not have been populated during acquisition, resulting in obsolete 
contents (see Fig.~\ref{fig:version} (left)). 
To resolve the conflict, the
client refetches the content of obsolete entries from MNs.

\stitle{Shared holders.}
In cases where a reader is enqueued when the queue contains
no writers, it becomes a \emph{shared holder}
without populating its queue entry, making the entry
obsolete (see Fig.~\ref{fig:version} (right)). 
Leveraging the fact that writers waiting for the lock
always populate their entries, the CQL protocol identifies
shared holders in the queue by excluding writers from the queue.
Accordingly, when a reader releases the lock, it locates 
all writers in the queue immediately after fetching the queue. 
Since entries of writers might not be populated yet,
it repeatedly fetches the queue until the number of valid writers
in the queue matches \texttt{wcnt} in the header.
Once the position of writers is determined, the remaining 
entries, whether valid or obsolete, certainly correspond to readers. 
This enables the reader releasing the lock to determine whether
its successor is a writer (Lines 11--13 in Fig.~\ref{fig:qlockop}).

\subsection{Lock Reset}\label{sec:proto:reset}

{\sys} resets the CQL lock states when a waiter is unable to obtain lock 
ownership in finite time due to corrupted lock states or node failures.
The reset process is exclusively executed by the client
initiating it.
A lock is considered to be undergoing a reset when its header contains 
a non-zero reset ID. 
Any lock operations that encounter a non-zero reset ID will be aborted.
Applications must retry aborted acquisition operations  
and can safely ignore aborted release operations.

\stitle{Reset occasions.} Clients may wait infinitely for a CQL lock 
in the following cases, which are handled through resets.

\etitle{Queue entry overwrite.}
If the content of a queue entry is overwritten (e.g., due to lock queue
overflows) before it is fetched by the client releasing the lock, 
the original content is lost.
Overwrites are detected when the fetched queue entry version is
larger than that calculated using the queue index.
The overwrite can be avoided by adopting a lock queue capacity 
that exceeds the maximum number of clients.

\etitle{Version overflow.}
If the version value keeps increasing after experiencing an integer overflow,
it will eventually increase to the initial version of queue entries 
(\texttt{-1}). 
Since only waiters populate their queue entries, some entries may remain
unpopulated and retain their initial values,
which cannot be distinguished from valid entries with a version equal
to \texttt{-1}.
Using 16-bit versions, version overflow occurs 
less than once in every 100,000 acquisitions of the same lock.

\etitle{CN failure.}
Similar to spinlocks, CQL locks held by failed clients are never released,
blocking waiters infinitely. To ensure the liveness of locks,
{\sys} sets a timeout for acquisitions and resets locks that experience
timeouts.

\stitle{Reset process.}
The lock reset process consists of the following three steps executed
sequentially.

\etitle{Step 1: Set the reset ID.}
The client sets the reset ID to its CN ID using a CAS operation on the
lock header. In cases where the CAS operation fails due to concurrent 
lock header updates by other lock operations, it retries the operation.
During retries, if the client discovers that the reset ID is not zero,
signifying an ongoing reset process, it gives up the reset process to
ensure only one client executes the remaining steps.

\etitle{Step 2: Notify participants.}
The client broadcasts a reset signal to all clients and waits
until it has collected all responses.
Responses from failed clients are not awaited.
Clients holding the lock respond to the signal after releasing the lock,
whereas other clients respond immediately.
Clients waiting for the lock abort the acquisition operation after responding.

\etitle{Step 3: Reset the lock states.}
The client reinitializes the lock queue and resets the lock header value
with two WRITE operations issued in sequence, clearing the reset ID.

\stitle{Handling expired notifications.}
Notifications sent before the reset may arrive after the reset finishes, 
causing the lock to be incorrectly granted to subsequent waiters.
To handle expired notifications, {\sys} maintains a reset counter 
on all CNs for each lock. This counter is incremented by 1 when the lock is reset.
The reset signal broadcasted in Step 2 synchronizes the counter value.
Clients embed the local reset counter value in notifications
and ignore any notification with a reset count that is smaller than 
the local reset counter.

\subsection{Protocol Correctness}\label{sec:proto:proof}

We demonstrate the correctness of the CQL protocol by proving that it
preserves two key properties of locks: mutual exclusion and 
liveness~\cite{rwcc,rwmcs}, in the absence of failures.
Mutual exclusion ensures that all read-write and write-write conflicts are
resolved. Liveness guarantees that all clients obtain the lock ownership
within a finite time.
The full proof can be found in Appendix A (see supplementary material).

\subsection{Failure Handling}\label{sec:proto:failure}

{\sys} preserves the mutual exclusion property of locks when experiencing
CN, MN, and network failures, but it does not guarantee strong liveness
upon MN and network failures because locks become unavailable in these cases.
Fairness is not preserved in case of failures because clients may abort
ongoing lock operations and acquire the lock in a new order.
{\sys} uses a reliable coordinator to detect failures 
(e.g., using heartbeats~\cite{burrows2006chubby,zookeeper})
and handle network partitions (e.g., by shutting down nodes outside 
of the largest clique~\cite{voltdb,nifty}), which has been 
well-studied and is orthogonal to distributed locks~\cite{fisslock,dslr}.
Clients check the failure detection results
when waiting for the completion of RDMA operations 
or waiting for the responses of the reset signal. 
Aligning with state-of-the-art DM applications~\cite{smart,chime,motor},
{\sys} uses the Reliable Connection (RC) of RDMA to handle 
packet loss and re-ordering at the transport layer.
According to our experiments in \S\ref{sec:eval:ft},
{\sys} achieves strong liveness when experiencing CN failures,
and preserves correctness under MN failures.
Details about how failures are handled
in {\sys} can be found in Appendix B (see supplementary material).

%% file: design.tex
\section{Timestamp-based Hierarchical Locking}\label{sec:design}

\nospacestitle{Challenge: queue size variability.}
The CQL protocol uses a static lock queue capacity because implementing 
atomic circular queues with variable capacity is hard~\cite{circqueue}.
However, in DM applications, the number of clients contending for a 
lock can vary significantly due to the elasticity of compute resources.
Consequently, a small queue capacity can cause frequent overflows 
in high-contention scenarios, while a large queue capacity leads 
to the high consumption of both MN memory and MN-NIC bandwidth.

\stitle{Solution: timestamp (TS)-based hierarchical locking.}
We observe that DM applications tend to utilize CPU cores on the same 
CN to exploit data locality, rather than spreading clients across more 
CNs~\cite{smart,chime,sherman,ford}. Following this insight,
we address the challenge with a \emph{hierarchical} design that maintains
CQL lock waiters at the CN level and employs local locks on each CN to
resolve lock conflicts among its clients. 
This approach reduces the required queue capacity to the number
of CNs allocated to the application.

However, hierarchical designs often compromise fairness among clients
across different CNs~\cite{cohort}. Specifically, local waiters may
preempt remote waiters who acquired the lock earlier, which cannot be
addressed by simply limiting the maximum consecutive ownership
transfers between local clients~\cite{sherman,alock}.
We ensure fairness by embedding \emph{timestamps} of lock acquisition
in the lock queue entries, which help maintain the correct order
between local and remote waiters.

\subsection{Lock Data Structures}

In {\sys}, each lock consists of a CQL lock on the MN that stores the data 
and a local lock on each CN.

\begin{figure}[t]
  \vspace{1mm}
  \begin{minipage}{1.\linewidth}
    \centering\includegraphics[scale=.44]{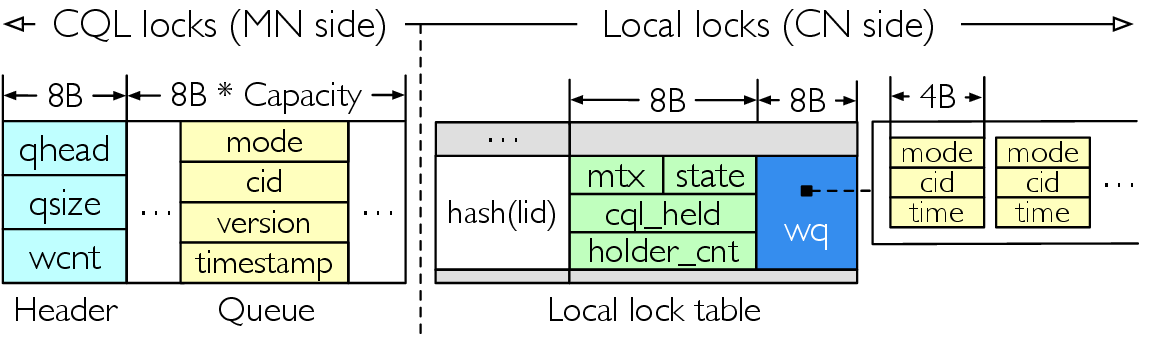}
  \end{minipage} \\[-2pt]
  \begin{minipage}{1.\linewidth}
    \caption{\small{\emph{Data structures of CQL locks (left) and local
    locks (right).
    }}}
    \label{fig:cohortds}
  \end{minipage}  \\[-15pt]
\end{figure}

\stitle{CQL locks.}
As shown in Fig.~\ref{fig:cohortds} (left), 
each CQL lock consists of an 8-byte header and several 8-byte queue entries. 
In addition to the mode, client ID, and version required by the CQL protocol,
each entry also records a timestamp indicating when the acquisition
operation is issued (see details in \S\ref{sec:design:ts}). 

\stitle{Local locks.}
As shown in Fig.~\ref{fig:cohortds} (right), each CN maintains a
hash table of local locks. Each local lock includes 
a mutex to protect its metadata (\texttt{mtx}), the local lock 
state (\texttt{state}), the holder count (\texttt{holder_cnt}), a flag 
indicating whether the CQL lock is held by the local CN (\texttt{cql_held}), 
and a pointer to a dynamically allocated queue of local waiters (\texttt{wq}). 
For each waiter, the queue records the acquisition mode (\texttt{mode}),
the client ID (\texttt{cid}), and an acquisition timestamp (\texttt{time}). 
When a lock is acquired, the corresponding local lock is created and 
inserted into the table if it is not already present. 
Throughout our experiments in \S\ref{sec:eval}, local lock tables
consume less than 20\,MB of memory on each CN, which is negligible
given that CNs typically have GBs of memory~\cite{fusee,smart}.

\subsection{Hierarchical Lock Operations}

\begin{figure}[t]
  \begin{minipage}{1.\linewidth}
    \centering\includegraphics[scale=.72]{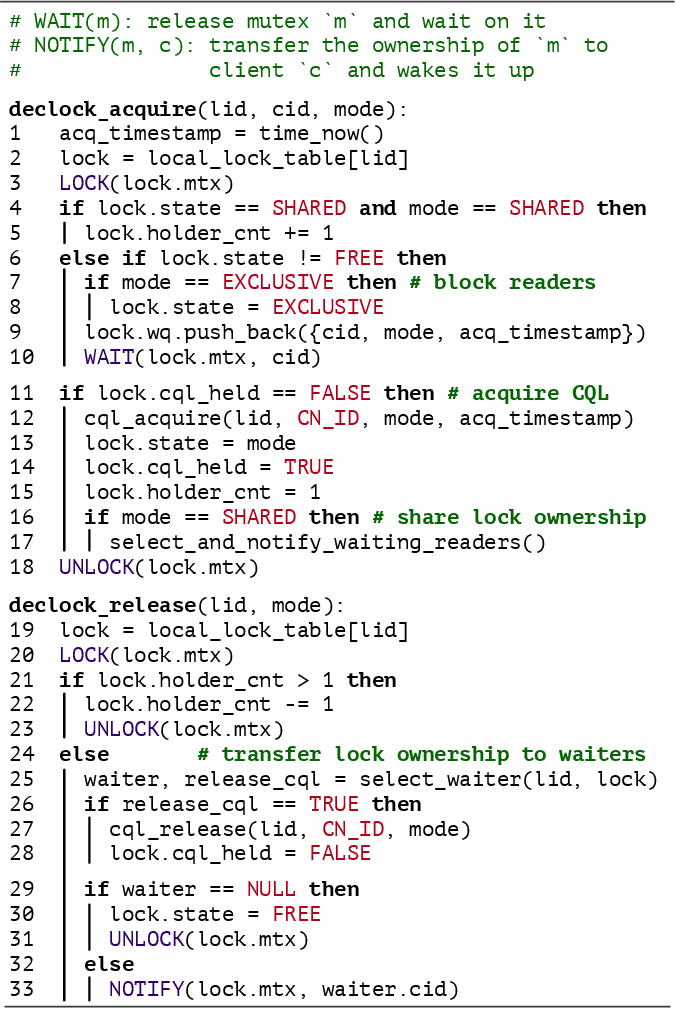}
  \end{minipage} \\[-10pt]
  \begin{minipage}{1.\linewidth}
    \caption{\small{\emph{Pseudocode of acquiring and releasing locks
    in {\sys}.
    }}}
    \label{fig:cohortop}
  \end{minipage}  \\[-15pt]
\end{figure}

Fig.~\ref{fig:cohortop} describes lock acquisition and release
operations in {\sys}. 
Fairness is mainly influenced by the waiter selection policy
(Lines 16 and 24), which is detailed in \S\ref{sec:design:ts}.

\stitle{Lock acquisition.} 
To acquire a lock, the client first locates the local lock in 
the table and acquires its mutex to gain exclusive access to the lock 
metadata (Lines 2--3). If a reader acquires a local lock in \texttt{SHARED} 
state, it immediately obtains lock ownership and increments the holder 
count (Lines 4--5). Otherwise, unless the lock state is \texttt{FREE}, 
the client appends itself to the wait queue and waits for notifications
from the current lock holders (Lines 9--10). 
Once a writer enters the wait queue, the lock state
transitions to \texttt{EXCLUSIVE}, which blocks subsequent readers 
to preserve the lock acquisition order (Lines 7--8).
After obtaining local lock ownership, if the CQL lock is not already 
held by the local CN, the client acquires the CQL lock
and updates the local lock metadata (Lines 12--15). 
If the client is a reader, it shares lock ownership
with selected readers in the wait queue by notifying them and
incrementing the holder count (Lines 16--17). 
Finally, the client releases the local lock's mutex (Line 18).

\stitle{Lock release.} 
To release a lock, the client first locates the local lock and
acquires its mutex, similar to the acquisition process (Lines 19--20).
If the holder count is greater than 1, indicating that lock
ownership is shared with other local clients, the client simply
decreases the holder count, releases the mutex, and returns
(Lines 21--23). Otherwise, the client selects a local waiter and
determines whether to transfer ownership directly to it based on
specific policies (Line 25).
If ownership should be transferred to remote waiters before
the selected local waiter, the client releases the CQL lock and clears
the \texttt{cql_held} flag, allowing the selected waiter to reacquire
the CQL lock later (Lines 26--28).
If there are no local waiters, the client sets the local lock 
state back to \texttt{FREE} and releases the mutex (Lines 29--31).
Otherwise, it notifies the selected local waiter (Lines 32--33). 

\subsection{Timestamp-based Fairness}\label{sec:design:ts}

\begin{figure}[t]
  \begin{minipage}{1.\linewidth}
    \centering\includegraphics[scale=.46]{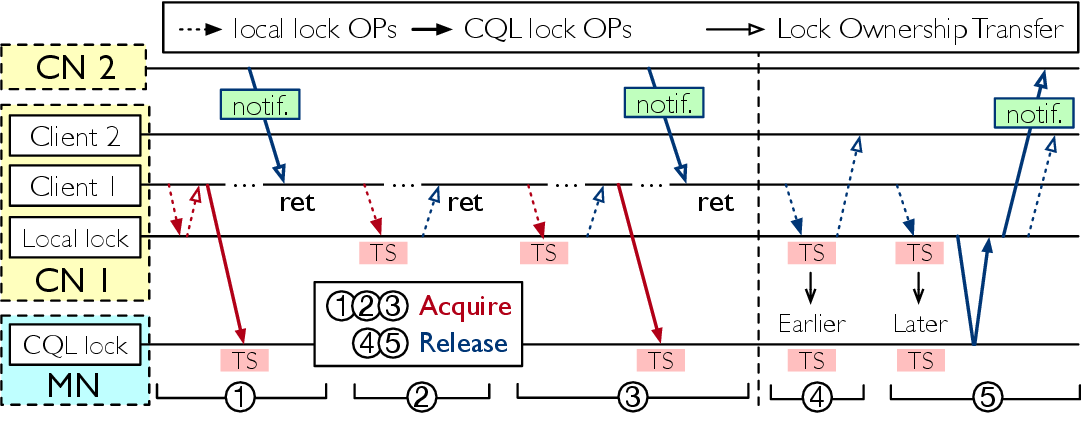}
  \end{minipage} \\[-2pt]
  \begin{minipage}{1.\linewidth}
    \caption{\small{\emph{Acquisition and release workflows of
    hierarchical locking where the acquisition timestamp (TS) 
    is recorded and compared.
    }}}
    \label{fig:timestamp}
  \end{minipage}  \\[-10pt]
\end{figure}

A key challenge in applying the hierarchical design in {\sys} is
maintaining fairness between local and remote waiters.
Fortunately, in CQL locks, waiter information is recorded in 
a centralized queue that is visible to clients across all CNs.
By embedding the acquisition timestamp in this information,
clients can determine the order of acquisition among waiters on
different CNs, thereby preserving fairness.

\stitle{Use cases of timestamps.}
Fig.~\ref{fig:timestamp} illustrates how the acquisition timestamp 
is recorded during lock acquisition
({\ding[1.2]{192}}--{\ding[1.2]{194}}) and compared
during lock release ({\ding[1.2]{195}}--{\ding[1.2]{196}}).
When a client needs to wait for the CQL lock
({\ding[1.2]{192}}/{\ding[1.2]{194}}),
it writes the timestamp along with other waiter information 
to the queue entry (Line 4 in Fig.~\ref{fig:qlockop}). 
Similarly, if the client needs to wait for the local lock 
({\ding[1.2]{193}}/{\ding[1.2]{194}}), 
it records the timestamp in its wait queue entry
(Line 8 in Fig.~\ref{fig:cohortop}). 
When releasing the lock, the client compares the timestamp of 
local and remote waiters (Line 24 in Fig.~\ref{fig:cohortop}).
If the first local waiter has an earlier timestamp than remote 
waiters, or if there are no remote waiters ({\ding[1.2]{195}}), 
it can directly obtain ownership without releasing the CQL lock
({\ding[1.2]{193}}). Otherwise ({\ding[1.2]{196}}), 
the client releases the CQL lock to transfer ownership
to remote waiters, requiring the local waiter to reacquire the 
CQL lock later ({\ding[1.2]{194}}).
Notably, the CQL lock must be reacquired to update
its state if the local waiter's acquisition mode differs from 
the local lock state.

\stitle{Fairness policies.}
{\sys} supports both task-fair and phase-fair policies to
accommodate different workloads~\cite{phasefair2}.
In the task-fair design, when a reader acquires the CQL lock 
and shares ownership with other waiting readers
(Lines 16--17 in Fig.~\ref{fig:cohortop}), 
it scans the wait queue to collect adjacent readers and
stops upon encountering a writer or a waiter with a later timestamp 
than remote waiters. In contrast, the phase-fair design allows the 
reader to share lock ownership with all readers in the wait queue.
Moreover, when a writer releases the CQL lock, the task-fair
design assigns the first local waiter to reacquire the CQL lock,
while the phase-fair design gives this privilege to the first reader
in the wait queue. In summary, the phase-fair design trades fairness
for higher concurrency by allowing waiting readers to preempt waiting writers.

\stitle{Synchronized time.}
{\sys} regularly synchronizes time on all CNs to enable
a direct comparison of acquisition timestamps between their clients.
This is achieved through
a counter on the MN. 
During synchronization, %
each CN atomically increments the counter by one using an FAA operation.
When the counter matches the total number of CNs, that CN resets it to 0.
All CNs repeatedly read the counter until it becomes 0
and subsequently record the current time.
This mechanism ensures precision within the latency of 
one RDMA READ (typically less than 3\,\us), which sufficiently ensures accurate 
ordering of lock acquisitions.
While our implementation uses a one-second synchronization interval,
this can be set smaller in systems with less stable clocks.

\stitle{Timestamp encoding.}
Each timestamp is a 16-bit unsigned integer recording the number of
microseconds passed since the time is synchronized. 
To address overflows, if the difference between two timestamps 
is greater than half of the maximum timestamp value (around 32\,ms), 
the timestamp with the larger value is considered to be the earlier timestamp. 
Notably, even if timestamp comparisons occasionally produce wrong results 
(e.g., when the difference is too large to be represented by 16 bits---which 
did not occur in our experiments), this only
affects fairness among the current waiters and does not compromise correctness.

\stitle{Prefetched remote timestamp.}
To check the timestamp of remote waiters, the client must fetch the remote 
lock queue, 
introducing extra latency to ownership transfer.
{\sys} %
hides this latency %
by delegating the task to clients waiting for the local lock. 
When a client starts waiting for the local lock 
(Line 10 in Fig.~\ref{fig:cohortop}),
it prefetches the remote lock queue and stores the earliest 
timestamp among remote waiters in the local lock. 
When the lock is released, the client 
can directly use the prefetched timestamp. %
If the timestamp is not prefetched, it indicates that either there are no 
remote waiters, or the remote waiter acquires the lock later than all 
local waiters. 
Thus, the lock ownership can be transferred to local waiters.

The prefetched remote timestamp becomes invalid when the CQL lock is released.
Therefore, clients must update the prefetched remote timestamp after
reacquiring the CQL lock (Line 12 in Fig.~\ref{fig:cohortop}).
To facilitate this, {\sys} requires clients releasing the CQL lock to embed 
the earliest timestamp from the queue in the notification, allowing the
waiter to update the prefetched timestamp without fetching the queue.

%% file: eval.tex
\section{Evaluation}
\label{sec:eval}

\subsection{Experimental Setup}

\nospacestitle{Testbed.}
The experiments were conducted on five machines in a single rack, each equipped
with two 24-core Intel CPUs, 128\,GB of RAM, and two ConnectX-4 100\,Gbps RDMA NICs. 
All RNICs are connected to a Mellanox 100\,Gbps switch, forming a unified
network.\footnote{\footnotesize{The performance results in a heterogeneous 
network can be found in Appendix C (see supplementary material).}}
Each machine hosts two logical nodes, each has one CPU connected with one RNIC 
used by all cores on it. We employ 8 nodes as CNs and 1 node as the
MN, aligning with prior work~\cite{ditto,smart} in the CN-MN ratio. 
Each CN dedicates 16 cores to running application clients.
Each core runs multiple clients as coroutines.

\stitle{Comparing targets.}
We compare {\sys} with three RDMA-based locks: CAS-based 
spinlock ({CASLock}), the improved version of DSLR~\cite{dslr}
({DSLR+} as mentioned in \S\ref{sec:bg:prior}), and 
ShiftLock~\cite{shiftlock}. CASLock represents the 
conventional spinlock mechanism commonly used in existing DM 
applications~\cite{ford,sherman,smart,clover}. 
For a fair comparison, we use a recent implementation with 
reader-writer semantics~\cite{rdmarw}.
DSLR+ combines the ticket lock algorithm~\cite{bakery} with
truncated exponential backoff~\cite{smart24} to optimize spinlocks. 
ShiftLock is a concurrent work with {\sys} that implements a 
reader-writer version of the MCS lock algorithm~\cite{mcs}
to reduce retries in spinlocks.
In all experiments, we set the lock queue capacity of 
{\sys} to the number of CNs (8 by default) and 
use 16-bit versions in lock queue entries.

\stitle{Applications.}
We build a microbenchmark to evaluate the performance of DM lock mechanisms
under various workload patterns.
The benchmark simulates typical DM lock usage~\cite{sherman,smart,ford,chime},
where each operation acquires 
a lock in shared (or exclusive) mode, fetches (or updates) the 
corresponding remote object,
and releases the lock. 
Furthermore, to demonstrate end-to-end benefits for DM applications, 
we integrate {\sys} into two DM applications (an object store and
a database index~\cite{sherman}) and evaluate them using 
real-world traces~\cite{twittertrace} and representative 
workloads~\cite{sherman}.

\begin{figure}[t]
  \vspace{0mm}
  \begin{minipage}{1.\linewidth}
    \centering\includegraphics[scale=.33]{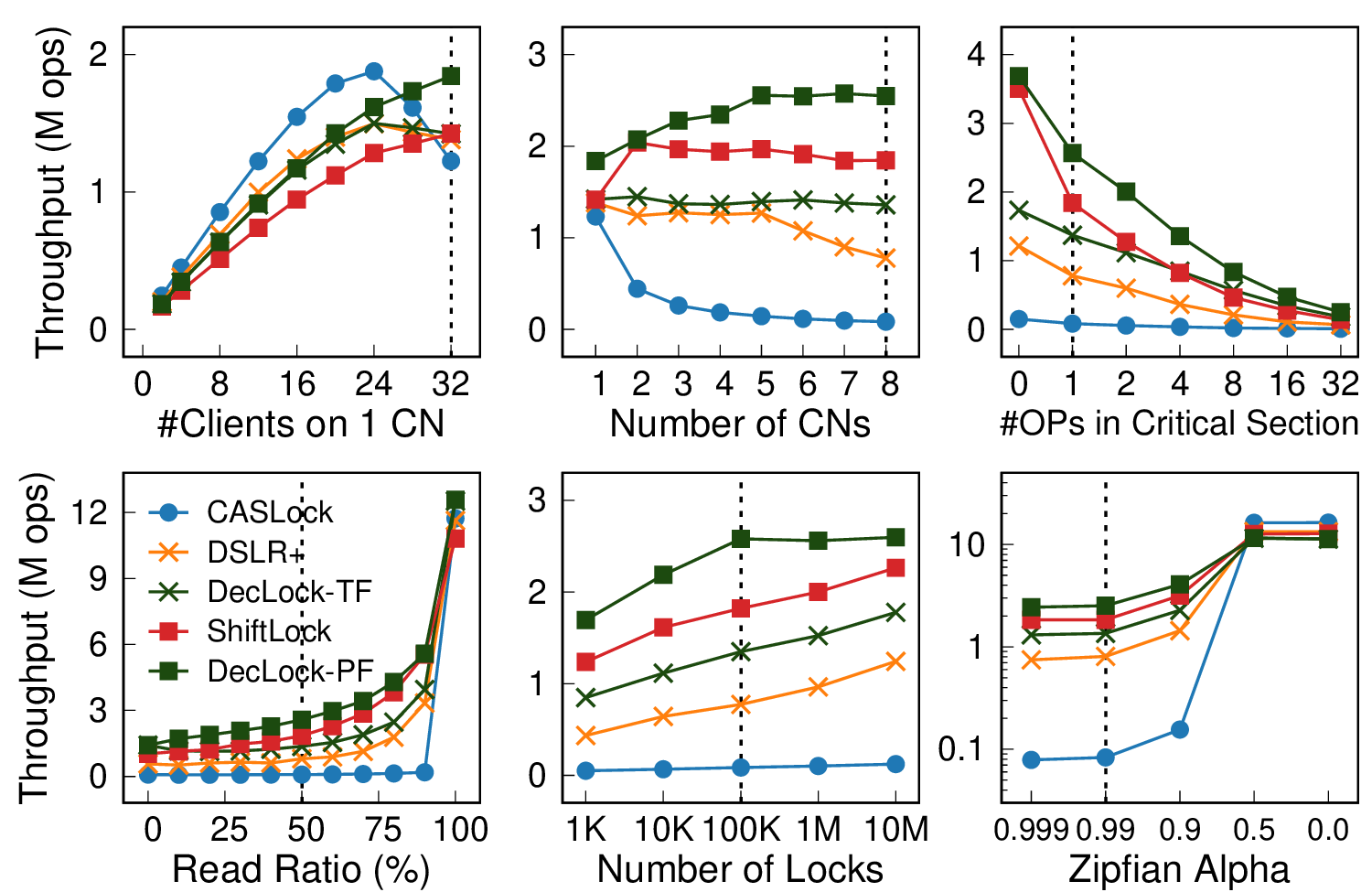}
  \end{minipage} \\[0pt]
  \begin{minipage}{1.\linewidth}
    \caption{\small{\emph{The operation throughput of the 
    microbenchmark when using different lock mechanisms 
    with different workload parameters.
    Dotted lines denote the default parameters.
    }}}
    \label{fig:lock-thpt}
  \end{minipage}  \\[-10pt]
\end{figure}

\subsection{Lock Operation Performance}\label{sec:eval:micro}

We first study the lock operation performance of {\sys} with 
the microbenchmark. 
{\sys} is evaluated with both task-fair (TF) and phase-fair (PF) 
policies, which match the fairness levels of DSLR+ and ShiftLock,
respectively.

\stitle{Throughput.}
We run the microbenchmark with different parameters to simulate various 
workload patterns and examine their impact on lock throughput.
By default, we use 32 clients on each of 8 CNs (256 in total),
with each client accessing 100,000 locks and corresponding 
objects following a Zipfian distribution with $\alpha=0.99$
and a 50\% read ratio.
These parameters reflect the access patterns of DM
applications under write-intensive workloads~\cite{sherman,smart}.
Fig.~\ref{fig:lock-thpt} shows the throughput of the microbenchmark
with different parameters.

\etitle{Number of clients.}
We evaluate scalability by increasing the number of clients. We start
with a single CN running different numbers of clients,
then scale out to multiple CNs. CASLock achieves the highest throughput
with up to 24 clients due to its simple acquisition logic, outperforming 
{\sys}, DSLR+, and ShiftLock by 1.25\x, 1.26\x, and 1.46\x, respectively. 
However, its throughput rapidly collapses beyond this point as 
the MN-NIC becomes overwhelmed by retries. 

\begin{figure}[t]
  \begin{minipage}{1.\linewidth}
    \centering\includegraphics[scale=.33]{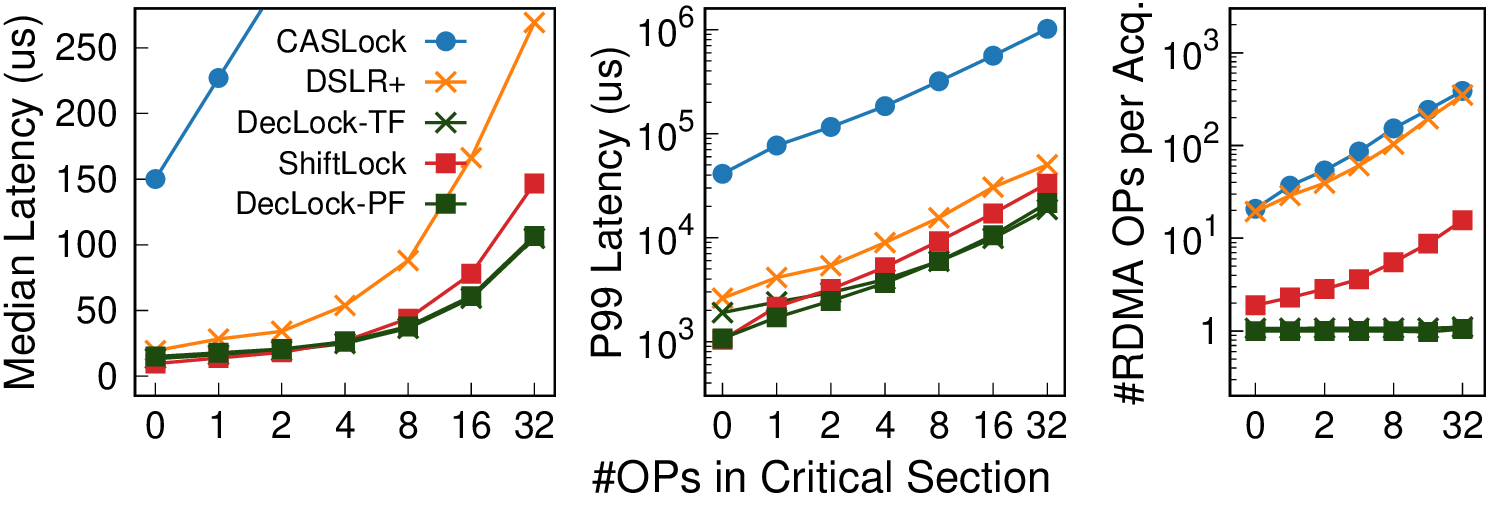}
  \end{minipage} \\[-5pt]
  \begin{minipage}{1.\linewidth}
    \caption{\small{\emph{The median (left) and \p{99} (middle) 
    operation latency and the average number of RDMA operations 
    per acquisition (right) of lock mechanisms under
    the microbenchmark.
    }}}
    \label{fig:latency}
  \end{minipage}  \\[-10pt]
\end{figure}

Although DSLR+ and ShiftLock reduce
MN-NIC IOPS usage through techniques such as backoff and direct handover,
their reliance on repeatedly reading lock states still leads to
MN-NIC contention, %
causing throughput degradation by 38.9\% and 9.5\%, respectively, 
as the number of CNs increases. 
In contrast, {\sys} maintains stable throughput after reaching its 
peak because it requires only one RDMA operation for lock acquisition
on average (see Fig.~\ref{fig:latency} (right)). 
Consequently, while achieving comparable 
fairness, {\sys}-TF outperforms DSLR+ by up to 1.74\x,
and {\sys}-PF outperforms ShiftLock by up to 1.40\x.

\etitle{Length of critical sections.}
We vary the number of RDMA operations that access remote objects
after acquiring each lock to simulate workloads with different
critical section lengths. As the length of critical section increases,
all lock mechanisms show decreasing throughput due to 
intensified lock conflicts. The longer wait time caused by these 
conflicts leads to higher average numbers of retries when acquiring CASLock
(387.9), DSLR+ (349.2), and ShiftLock (15.6), 
as shown in Fig.~\ref{fig:latency} (right). In contrast, {\sys} 
requires at most 1.10 RDMA operations on average for acquisition, 
regardless of the critical section length, as it eliminates
retries through fully decentralized ownership transfer.
The extra 0.10 operations come from filling the queue entry when
waiting for lock ownership.
This significant reduction in MN-NIC IOPS consumption leads to 
throughput improvements over CASLock (up to 43.47\x), DSLR+ (up to 4.35\x),
and ShiftLock (up to 1.81\x).

\etitle{Read ratio.}
{\sys}-TF consistently outperforms DSLR+ (by 1.18\x--2.50\x) and CASLock 
(by 15.31\x--20.80\x) across all read ratios except 100\%, due to reduced 
MN-NIC IOPS consumption. ShiftLock exceeds {\sys}-TF by up to 1.55\x because
its writers occasionally transfer ownership to all waiting readers, 
sacrificing fairness to improve concurrency. {\sys}-PF, which uses a similar 
strategy among local waiters, outperforms ShiftLock by 1.12\x--1.55\x. 
All lock mechanisms have similar performance in read-only scenarios, as all 
acquisitions obtain the lock without retries.

\etitle{Number of locks.}
As the number of locks increases from 1,000 to 10 million, 
the throughput of lock mechanisms rises by 2.49\x (CASLock),
2.86\x (DSLR+), 1.83\x (ShiftLock), 2.11\x ({\sys}-TF), and 1.54\x 
({\sys}-PF). This improvement occurs because lock acquisitions are 
less likely to conflict with a larger range of locks. 
However, {\sys}-PF's throughput stops 
growing after the number of locks exceeds 100,000, as the higher 
lock count also reduces the chance of transferring lock ownership 
among local clients.

\etitle{Skewness.}
We decrease the $\alpha$ parameter of the Zipfian distribution to reduce
workload skewness, where $\alpha=0$ indicates a uniform distribution.
As the workload becomes less skewed, the throughput of CASLock, DSLR+, 
and ShiftLock converges to that of {\sys}, since the number of retries 
approaches zero due to fewer lock conflicts.

\begin{figure}[t]
  \begin{minipage}{1.\linewidth}
    \centering\includegraphics[scale=.33]{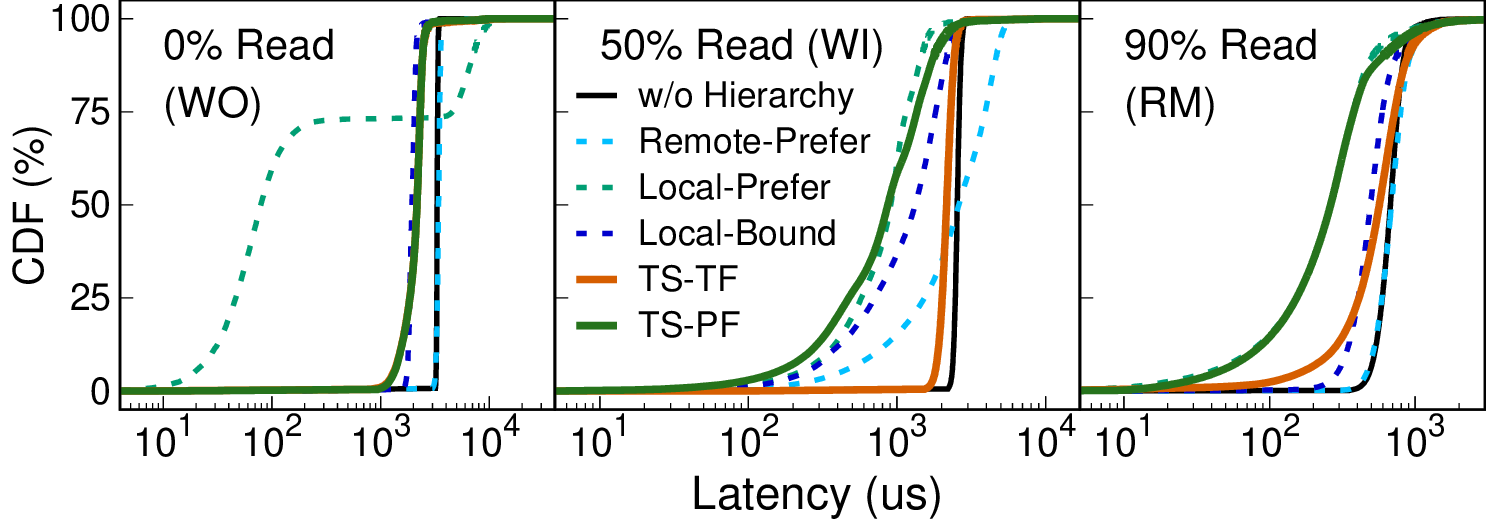}
  \end{minipage} \\[-3pt]
  \begin{minipage}{1.\linewidth}
    \caption{\small{\emph{The acquisition latency distribution
      of the most contended lock in the microbenchmark
      with different hierarchical designs.
    }}}
    \label{fig:fairness}
  \end{minipage}  \\[-13pt]
\end{figure}

\stitle{Latency.}
We investigate how lock mechanisms affect the 
end-to-end latency of microbenchmark operations, as shown in
Fig.~\ref{fig:latency}. Due to space constraints, our analysis 
focuses on operation latency across different
critical section lengths.%

\etitle{Median latency.}
The median operation latency reflects cases where the lock is
acquired immediately, without retries. {\sys}'s median latency is
proportional to the critical section length, as its acquisition
and release latencies remain constant. In contrast, other lock 
mechanisms experience rapidly increasing latency, leading to a widening
gap compared to {\sys}. This is because the MN-NIC IOPS is 
saturated by retries for a few hot locks, which slows down  
all RDMA operations, including lock acquisitions without retries.
Despite a similar number of retries, CASLock
has significantly higher latency than DSLR+ because it uses
CAS for retries, which consumes much more IOPS than READ. 
In summary, {\sys} reduces the median
latency of CASLock, DSLR+, and ShiftLock by up to 95.8\%,
64.3\%, and 28.5\%, respectively.

\etitle{Tail (\p{99}) latency.}
The tail operation latency reflects the time required to acquire 
heavily contended locks.
CASLock exhibits over an order of magnitude higher tail latency than
other lock mechanisms due to its lack of fairness. DSLR+ and ShiftLock
also have higher tail latency than {\sys} because they experience greater
RDMA operation latency and spend more time executing critical
sections, which increases the aggregated wait time. Initially,
{\sys}-PF has lower tail latency than {\sys}-TF due to
higher concurrency. However, as the critical section length increases,
{\sys}-PF's tail latency surpasses that of {\sys}-TF because of 
weaker fairness. In summary, {\sys} reduces the \p{99} latency of 
CASLock, DSLR+, and ShiftLock by up to 98.2\%,
67.8\%, and 43.7\%, respectively.

\subsection{Hierarchical Designs}

We demonstrate the efficiency and fairness of the timestamp mechanism 
by comparing it with other ownership transfer policies in hierarchical 
locking. These policies include prioritizing remote waiters over local 
ones (Remote-Prefer), prioritizing local waiters over remote ones 
(Local-Prefer), and transferring ownership to remote waiters after at 
most $N$ consecutive local transfers~\cite{alock,sherman} (Local-Bound), 
where we set $N=4$ to align with prior work~\cite{sherman}.
All three policies use the phase-fair design between local 
readers and writers, while the timestamp mechanism is evaluated
with both task-fair (TS-TF) and phase-fair (TS-PF) designs.
We run the microbenchmark with three read ratios representing
write-only (WO, 0\%), write-intensive (WI, 50\%), and read-mostly 
(RM, 90\%) workloads and analyze the latency distribution of the
most contended lock to highlight the impact of fairness.

\begin{figure}[t]
  \begin{minipage}{1.\linewidth}
    \centering\includegraphics[scale=.33]{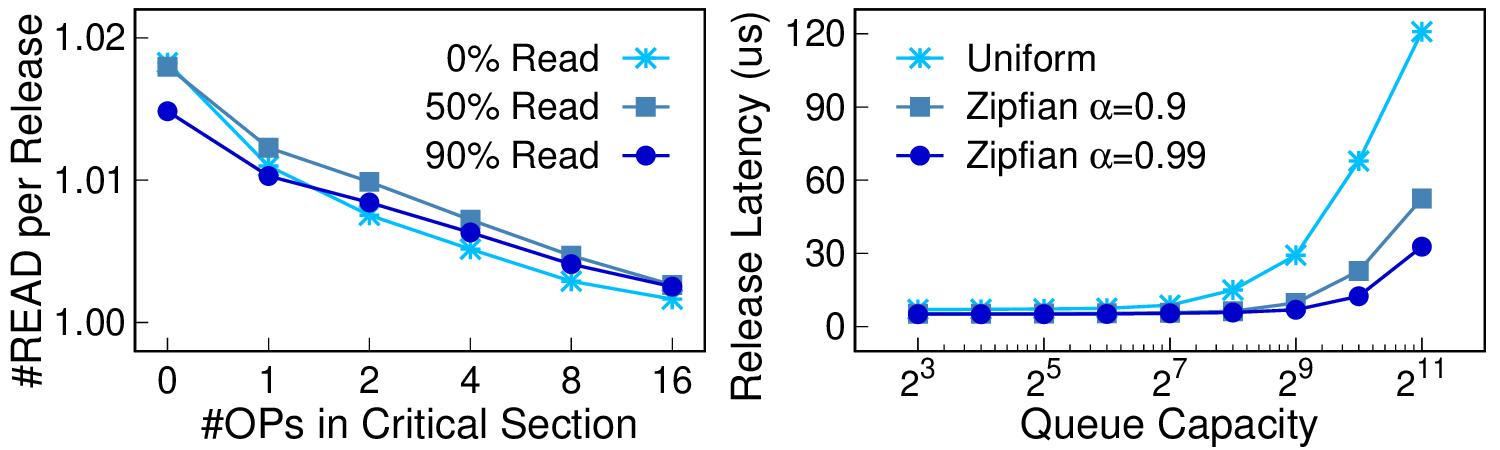}
  \end{minipage} \\[-3pt]
  \begin{minipage}{1.\linewidth}
    \caption{\small{\emph{The average number of extra RDMA READs per lock
    release caused by refetching obsolete entries (left) and the median 
    lock release latency with different capacities of the lock queue (right).
    }}}
    \label{fig:factor}
  \end{minipage}  \\[-10pt]
\end{figure}

As shown in Fig.~\ref{fig:fairness}, in WO and RM workloads, 
the remote-prefer policy exhibits a latency distribution similar 
to that without hierarchical locking, as its waiters always 
reacquire the remote CQL lock after being notified. However, 
in WI workloads, it shows higher latency variance because 
the phase-fair design prioritizes readers, increasing writers' 
wait times. In contrast, the local-prefer policy delivers 
performance comparable to TS-PF in WI and RM workloads
by maximizing reader concurrency.
However, it can starve remote waiters, resulting in significant
tail latency in WO workloads.
The local-bound policy closely matches TS-PF's performance in 
WO workloads, but its latency gap with TS-PF widens
as the read ratio increases because it allows only a fixed number
of local readers to share the lock, limiting concurrency.

Across all read ratios, the timestamp mechanism achieves lower
median and \p{99} latencies compared to not using the hierarchical 
design because it accelerates ownership transfer among local waiters
while preserving fairness across CNs. Compared with TS-TF, TS-PF
further reduces latency in WI and RM workloads by allowing 
more readers to share the lock without being blocked by waiting
writers. Overall, compared to other ownership transfer policies, 
the timestamp mechanism reduces the median and \p{99} acquisition 
latencies by up to 66.7\% and 70.8\%, respectively.

\subsection{Entry Refetching Overhead}\label{sec:eval:refetch}

We measure the additional RDMA READ operations required to handle 
obsolete entries (see \S\ref{sec:proto:version}) under various
microbenchmark parameters, as shown in Fig.~\ref{fig:factor} (left).
In all configurations, {\sys} incurs an average maximum of 
0.018 extra RDMA operations per release, which consumes negligible
MN-NIC resources. Moreover, the refetching frequency is inversely 
proportional to the length of critical sections because when clients 
hold the lock for longer periods, waiters have more time to update
their entries.

\subsection{Impact of Lock Queue Capacity}

We investigate the impact of lock queue capacity on {\sys} 
by measuring its release latency across various capacities, as
shown in Fig.~\ref{fig:factor} (right). 
As capacity increases, release latency rises more quickly in workloads
with lower skewness because these workloads have higher operation throughput 
and saturate MN bandwidth earlier. Nevertheless, release latency
remains stable below 8.80\,\us when the queue capacity is up to 128, 
even in uniform workloads. This result indicates that {\sys} can support 
128 CNs with little performance loss. Notably,
this capacity far exceeds the typical number of CNs
assigned to a single DM application~\cite{polardbmp,chime,smart,motor}.

\begin{figure}[t]
  \begin{minipage}{1.\linewidth}
    \centering\includegraphics[scale=.33]{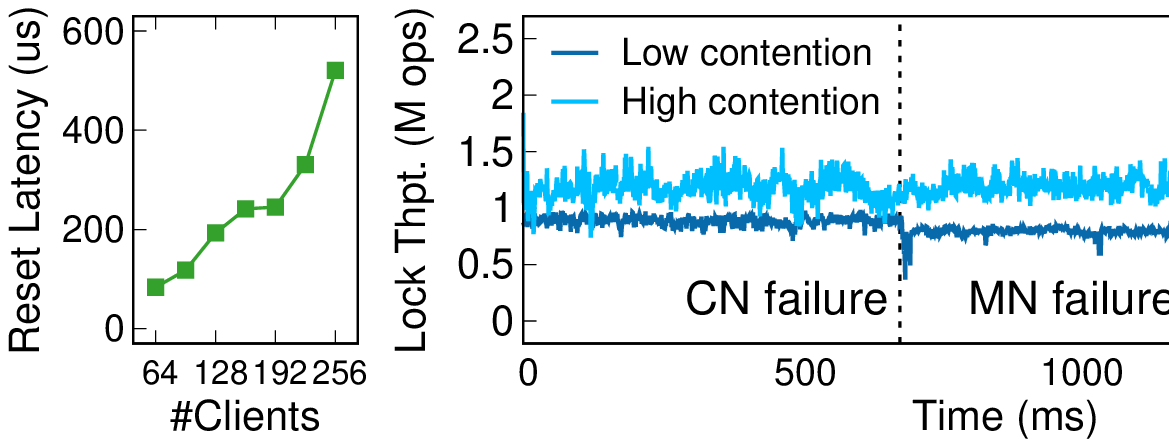}
  \end{minipage} \\[-2pt]
  \begin{minipage}{1.\linewidth}
    \caption{\small{\emph{The latency of resetting a lock with different 
    number of clients (left) and the timeline of {\sys}'s lock operation 
    throughput with the presence of CN and MN failures (right).
    }}}
    \label{fig:failure}
  \end{minipage}  \\[-10pt]
\end{figure}

\subsection{Lock Reset Overhead}

We measure the average reset latency with varying numbers of clients,
as shown in Fig.~\ref{fig:failure} (left). 
The average lock reset latency increases from 83.29\,\us to 520.72\,\us as 
the number of clients grows. This latency escalation primarily stems from 
the process of sending reset signals to clients and awaiting their responses.
Despite their relatively high latency, lock resets have minimal impact on 
the performance of {\sys} because of their infrequent occurrences.
{\sys} uses sufficiently large 
lock queues to prevent queue overflows and employs 16-bit versions to 
reduce version overflows. Consequently, across all experiments conducted in 
\S\ref{sec:eval:micro}, at most 0.0014\% of acquisitions are reset.

\subsection{Fault Tolerance}\label{sec:eval:ft}

To evaluate {\sys}'s liveness and correctness when tolerating CN and MN failures,
we manually shut down 1 CN and 1 MN in sequence while running the 
microbenchmark.
We conduct the workload in two configurations: one using 1 core 
({Low contention}) and the other using 16 cores ({High contention}) 
on each of 8 CNs.
As shown in Fig.~\ref{fig:failure} (right), under low contention,
the lock throughput quickly stabilizes at around 88\% of the original
throughput after 1 CN fails as one-eighth of the clients are terminated.
Under high contention, the throughput remains unchanged
after the failure since {\sys} maintains its peak throughput even with fewer clients.
When the MN fails, all operations pause and the lock throughput
drops to 0\,ops until the MN recovers.

\subsection{DM Application Performance}

We evaluate {\sys} using two DM applications to
demonstrate its effectiveness in mitigating the performance degradation
under high contention caused by locking. {\sys} uses the phase-fair
design in both applications.

\stitle{Object store.}
We build a DM object store that allows clients on CNs to perform get and 
set operations on objects stored on MNs. Accesses to objects are protected 
by reader-writer locks on MNs. Each client performs operations based on
object size, get-set ratio, and key distribution derived from real-world 
traces of Twitter~\cite{twittertrace}. 
We evaluate the object store using two traces: one with a small average
object size (414\,B) and a low get ratio (65\%) to represent IOPS-bound 
scenarios, and another with a large average object size (9213\,B) and
a high get ratio (89\%) to represent bandwidth-bound scenarios.

\begin{figure}[t]
  \vspace{1mm}
  \begin{minipage}{1.\linewidth}
    \centering\includegraphics[scale=.28]{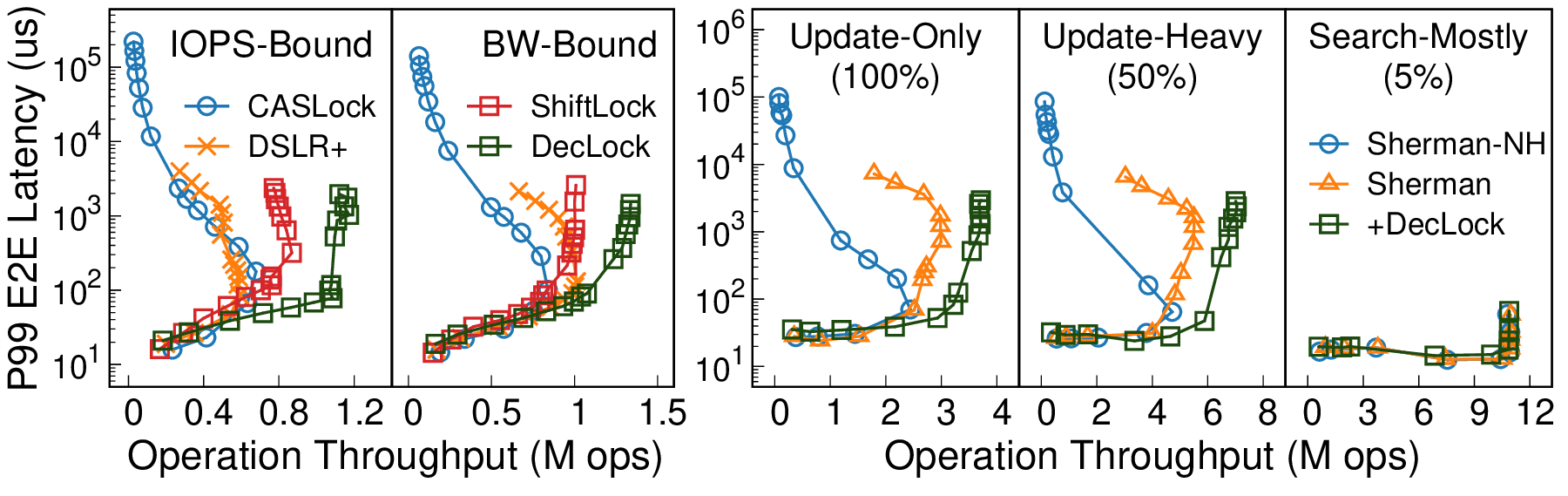}
  \end{minipage} \\[-5pt]
  \begin{minipage}{1.\linewidth}
    \caption{\small{\emph{The object store's throughput and \p{99}
      operation latency in IOPS-bound and 
      bandwidth (BW)-bound scenarios
      when using different lock mechanisms (left), and
      the update/search operation throughput and
    \p{99} operation latency of Sherman using different lock mechanisms
    under different workloads (right).
    }}}
    \label{fig:sherman}
  \end{minipage}  \\[-10pt]
\end{figure}

As shown in Fig.~\ref{fig:sherman} (left), using CASLock or DSLR+ causes
a throughput collapse in both scenarios because RDMA retries consume most 
of the MN-NIC IOPS, increasing the latency of other RDMA operations. 
ShiftLock partially alleviates this issue by eliminating retries from 
waiting writers. However, retries from waiting readers still consume 
most of the MN-NIC IOPS, limiting peak throughput. Moreover, its weaker
fairness results in higher tail latency than DSLR+ in bandwidth-bound
scenarios.
In contrast, {\sys} eliminates retries entirely, allowing MN-NICs to
operate efficiently and maintain stable throughput after reaching
IOPS or bandwidth bottlenecks. {\sys} improves the object store's 
throughput by up to 35.60\x over CASLock, 3.33\x over DSLR+, and 
1.48\x over ShiftLock. It also reduces the \p{99} latency by up to 
98.8\%, 30.3\%, and 79.7\% compared to CASLock, DSLR+, and ShiftLock,
respectively.

\stitle{Database index.}
Sherman~\cite{sherman} is a state-of-the-art B$^+$Tree index for DM.
It uses CAS-based locks to serialize tree node updates and
adopts hierarchical locking to reduce the cumulative 
rate of RDMA retries. 
We integrate {\sys} into Sherman and evaluate it against vanilla 
Sherman using the workloads from Sherman's paper~\cite{sherman},
including Update-Only (100\% updates), Update-Heavy 
(50\% updates and 50\% searches), 
and Search-Mostly (5\% updates and 95\% searches).
We also evaluate Sherman without hierarchical locking (Sherman-NH) 
as a baseline. Since Sherman does not support other lock mechanisms
like DSLR+ and ShiftLock, we do not evaluate them due to the significant
complexity of integration.
All systems use up to 512 clients
distributed across up to 32 CNs.\footnote{\footnotesize{We virtualize
each node into 4 CNs after the number of CNs exceeds 8.}}

As shown in Fig.~\ref{fig:sherman} (right), {\sys} improves throughput
by up to 27.71\x compared to Sherman-NH and 2.31\x compared to Sherman.
It also reduces the \p{99} latency by up to 97.6\% and 82.1\%, respectively.
Under Update-Only and Update-Heavy workloads,
Sherman-NH experiences a throughput collapse after 
reaching its peak because RDMA retries occupies the MN-NIC.
While hierarchical locking reduces retry frequency because only 
one client on each CN acquires the remote CASLock,
MN-NIC usage still increases as more CNs are added, eventually 
causing throughput collapse beyond 16 CNs.
In contrast, {\sys} enables Sherman to sustain high throughput and 
moderate tail latency with more clients by using MN-NICs efficiently.
Under Search-Mostly workload, all three lock mechanisms perform similarly 
since search operations in Sherman are lock-free.

%% file: related.tex
\section{Related Work}

\nospacestitle{Disaggregated memory applications.}
Numerous applications have been redesigned for DM 
to leverage its elasticity in compute and memory resources.
These applications include
hash tables~\cite{clover,dinomo,fusee,ditto,race}, 
transaction processing systems~\cite{ford,rtx,motor}, 
tree-based indexes~\cite{sherman,smart,rolex,marlin,chime}, 
and database systems~\cite{dsmdb,polardbsrvless,polardbmp}.
Many of them use CAS-based spinlocks to serialize concurrent data 
accesses~\cite{ford,motor,sherman,smart,rolex,chime,clover,rtx,marlin,dsmdb}, 
which can hamper performance under high contention due to inefficient 
MN-NIC IOPS usage and lack of fairness. 
{\sys} mitigates the performance implications of locking in DM 
applications through decentralized coordination, minimizing MN-NIC 
usage and ensuring fairness.

\stitle{RDMA-based distributed locks.}
Several studies have explored the design of distributed locks 
based solely on one-sided RDMA 
operations~\cite{ddlm,drtm,dslr,calvin,rdmarw,shiftlock,alock}.
Some adopt advanced lock algorithms, such as ticket
locks~\cite{dslr} and MCS locks~\cite{ddlm,alock,shiftlock},
which improve scalability and fairness over spinlocks.
Although ticket locks support reader-writer semantics with strict 
fairness, they still rely on repeated checks of lock states on MNs, 
failing to address contention on MN-NICs. MCS locks either 
do not support readers~\cite{ddlm,alock} or track readers using
counters~\cite{shiftlock}, which leads to inefficient use of 
MN-NICs and fairness violations between readers and writers.
By ordering all waiters in a centralized queue, {\sys} ensures
fairness and efficient MN-NIC usage, completely 
eliminating retries during lock acquisitions.

Inspired by the lock cohorting concept in NUMA systems~\cite{cohort},
some RDMA locks use hierarchical designs to improve 
scalability~\cite{alock,sherman}. However, these designs compromise
fairness across CNs, as they cannot track the acquisition order
among local and remote waiters. Additionally, they do not support
reader-writer semantics, limiting concurrency. In contrast, 
by recording the acquisition timestamp of remote waiters in 
centralized lock queues on MNs, {\sys} ensures fairness across CNs  
while supporting reader-writer semantics.

%% file: concl.tex
\section{Conclusion}

This paper presents {\sys}, a scalable, efficient and fair lock mechanism 
designed for disaggregated memory architecture. 
Evaluation using both microbenchmark and DM applications 
confirms the benefits of {\sys}.

%% file: appendix.tex
\begin{appendix}

\appendix
\renewcommand{\thesection}{Appendix \Alph{section}}

\section{Protocol Correctness}

We demonstrate the correctness of the CQL protocol by proving that it
preserves mutual exclusion and liveness properties in the absence of failures.
First, we describe and prove the correctness of the lock reset mechanism.
Then, we provide the proof for mutual exclusion and liveness properties.
We assume that the application always release locks it holds within a 
finite time, e.g., by setting a timeout for critical sections.
The protocol correctness upon failures is argued in the failure handling
subsection (\S\ref{sec:proto:failure}).

\subsection{Correctness of Lock Reset}

\stitle{Reset occasions.}
We first demonstrate that version overflows and queue entry overwrites 
can be detected in finite time, preventing waiters from waiting infinitely.
Version overflows are detected once the version calculated using the
lock header value reaches \texttt{-1}. Since this calculation occurs
during each release operation, and current holders release the lock in
finite time, version overflows are always detected promptly.
For queue entry overwrites, a waiter's entry is always examined when its 
predecessor releases the lock. Therefore, if a waiter cannot obtain lock
ownership because its entry has been overwritten, its predecessor is guaranteed
to detect this overflow when releasing the lock.

\stitle{Reset functionality.}
Next, we demonstrate that lock resets can correctly restore the initial 
state of a lock by showing that two invariants always hold:
(1) The reset process terminates in finite time. 
(2) The lock has no holder or waiter when the reset process terminates.

\vspace{5pt}
\stitle{Invariant 1.1}: \emph{The reset process terminates in finite time.}
\vspace{5pt}

\noindent
\emph{Proof:}
Each step in the reset process is finite. Step 1 might retry the CAS operation
on the lock header, but the number of retries is finite because concurrent
header updates are rare. Step 2 waits for responses from all surviving clients,
which is finite as all lock holders release the lock in finite time. 
Step 3 only takes 2 RDMA WRITEs and is therefore finite. 

\vspace{5pt}
\stitle{Invariant 1.2}: \emph{The lock has no holder or waiter when the 
reset process terminates.}
\vspace{5pt}

\noindent
\emph{Proof:}
Step 1 sets the reset ID, which prohibits new holders and waiters. 
In Step 2, all the existing holders release the lock before the reset process
can proceed, and all waiters abort the lock operation.
Therefore, the lock has no holder or waiter when the reset process 
terminates in Step 3.

\subsection{Proof of Mutual Exclusion}

For reader-writer locks, mutual exclusion guarantees that all reader-writer
and writer-writer conflicts are resolved by controlling the ownership of locks.
We show that CQL maintains mutual exclusion by proving Invariants 2.1 and 2.2.

\vspace{5pt}
\stitle{Invariant 2.1}: \emph{The ownership of locks held by a
writer can not be obtained by any other clients.}
\vspace{5pt}

\noindent
According to the lock acquisition workflow, the acquisition operation
can not obtain the lock ownership as long as the value of \texttt{wcnt}
field in the lock header is larger than 0 (Line 2). Hence, Invariant 1 holds
as long as the value of \texttt{wcnt} field is at least 1 when
the lock is held by a writer, which we prove as follows.

\etitle{Lemma 1}: \emph{The value of \texttt{wcnt} field in the
lock header is at least 1 when the lock is held by a writer.}
\vspace{5pt}

\noindent
\emph{Proof:}
When a client acquires the lock in exclusive mode, it atomically increases
the value of \texttt{wcnt} by 1 (Line 1). It does not change the value of
\texttt{wcnt} until it releases the lock, in which case it atomically 
decreases the value by 1 (Line 6). Therefore, the value of \texttt{wcnt} 
is consistent
with the number of writers holding the lock or waiting
for the lock. Hence, if the lock is being held by a writer,
the value of \texttt{wcnt} is at least 1.

\vspace{5pt}
\stitle{Invariant 2.2}: \emph{The ownership of locks held by readers 
can not be obtained by writers.}
\vspace{5pt}

\noindent
According to the lock acquisition workflow, an exclusive acquisition operation
can not obtain the lock ownership as long as the value of \texttt{qsize}
field in the lock header is larger than 0 (Line 2). 
Since all clients atomically increment the \texttt{qsize} when acquiring the 
lock and do not decrement it until the lock is released, the \texttt{qsize}
field is definitely larger than 0 if the lock is held by readers,
which prevents the ownership from being obtained by any writer.

\subsection{Proof of Liveness}

CQL guarantees that, as long as all application clients release the held locks
in a finite time, all locks can be obtained in a finite time.
We prove this liveness property as follows.

\vspace{5pt}
\stitle{Invariant 3}: \emph{All clients obtain the lock
ownership in a finite time as long as all holders release the
lock in a finite time.}
\vspace{5pt}

\noindent
According to the lock acquisition workflow, clients in case {\ding[1.2]{192}}
obtain the lock ownership immediately. Thus, we focus on discussing clients
in case {\ding[1.2]{193}} that need to wait for the lock ownership.
Given that the client always populates its queue entry in case 
{\ding[1.2]{193}}, and queue entry overwrites are prevented by lock resets, 
the waiter's information remains in the queue until the waiter is notified.
Hence, the waiting client can obtain the lock ownership in a finite time
as long as
all waiters recorded in the queue are notified in a finite time, which
we prove as follows.

\etitle{Lemma 2}: \emph{All waiters recorded in the queue are notified in a 
finite time.}
\vspace{5pt}

\noindent
\emph{Proof:}
When there are waiters in the lock queue, the \texttt{qsize} is always
larger than 1, and the \texttt{wcnt} is always larger than 0.
Otherwise, all clients in the lock queue fall in case {\ding[1.2]{192}},
i.e., they have obtained the lock ownership.
Therefore, as long as there are waiters in the queue, the client holding
the lock always notifies at least one waiter when releasing the lock
(cases {\ding[1.2]{195}}/{\ding[1.2]{196}}), which becomes the new lock
holder.
Hence, as long as all lock holders release the lock in a finite time,
all waiters recorded in the queue will eventually be notified.

\section{Failure Handling}

\stitle{Failure model.}
{\sys} preserves the mutual exclusion property of locks when experiencing
CN, MN, and network failures, but it does not guarantee strong liveness
upon MN and network failures because locks become unavailable in these cases.
Fairness is not preserved in case of failures because clients may abort
ongoing lock operations and acquire the lock in a new order.
{\sys} uses a reliable coordinator to detect failures 
(e.g., using heartbeats) and handle network partitions.
% (e.g., by shutting down nodes outside of
% the largest clique~\cite{voltdb}).
% or re-routing packets through intermediate
% nodes~\cite{nifty}.
Clients check the failure detection results
when waiting for the completion of RDMA operations 
% during lock operations (Lines 1 and 6 in Fig.~\ref{fig:qlockop}) 
or waiting for the responses of the reset signal. 
% (\S\ref{sec:proto:reset}).

\stitle{CN failures.}
If a CN fails, the locks held by clients on it will never be 
released, and notifications sent from or to it are lost. 
These consequences do not violate mutual exclusion, but harms
the liveness because the lock ownership can not be 
transferred to subsequent waiters. 
To maintain liveness, the first waiter who experiences the 
acquisition timeout resets the lock, aborting the acquisition of
other waiters (\S\ref{sec:proto:reset}).
Once the reset finishes, the waiters retry the acquisition,
excluding the influence of failed clients.

As a rare case, CN failures may occur during the reset process. 
If the client executing the reset process survives,
the reset process continues without obstruction from 
clients on failed CNs since their responses are not awaited.
If the client executing the reset process has failed, i.e., 
the reset ID corresponds to a failed CN, 
the surviving clients ignore reset signals from the failed CN
and attempt to initiate a new reset process. 
To achieve this, clients set the reset ID to their own CN ID with 
CAS operations, letting the first successful client execute the 
reset process.
This mechanism guarantees that the reset process always terminates
even when experiencing CN failures.

\stitle{MN failures.}
All lock operations towards failed MNs are paused until the MNs recover.
After the recovery, {\sys} aborts all paused operations and resets all 
locks on the failed MN before executing any new lock operation. 
This ensures that locks on failed MNs are in free states and have no actual
holders or waiters, preserving mutual exclusion.
% When a MN fails, the data and locks stored on it become temporarily
% inaccessible, and all RDMA operations towards the MN will experience 
% timeouts. {\sys} does not persist and restore lock states on MNs.
% Instead, after the MN is restarted, it aborts all lock operations 
% that experience RDMA timeout and reinitializes locks on the MN.

\stitle{Network failures.}
{\sys} uses the Reliable Connection (RC) of RDMA to handle 
packet loss and re-ordering at the transport layer.
{\sys} treats CNs or MNs that become unavailable due to network partitions
as failed ones. Upon complete partition, where nodes are split into disjoint 
groups, locks on MNs in each group are inaccessible from other groups.
Therefore, regarding clients in other groups as failed ones does not
violate mutual exclusion. When the partition finishes, locks on all MNs 
will be reset, converging their state to not having any holders or waiters. 
Upon partial partition, {\sys} requires the coordinator to guarantee that 
all nodes have a consistent view of failed nodes, such as by finding a 
maximum clique and shutting down other nodes. 
This allows {\sys} to handle the partition using the same approach as
handling node failures.

\section{Heterogeneous Network}

\begin{figure}[t]
  \vspace{2mm}
  \begin{minipage}{1.\linewidth}
    \centering\includegraphics[scale=.33]{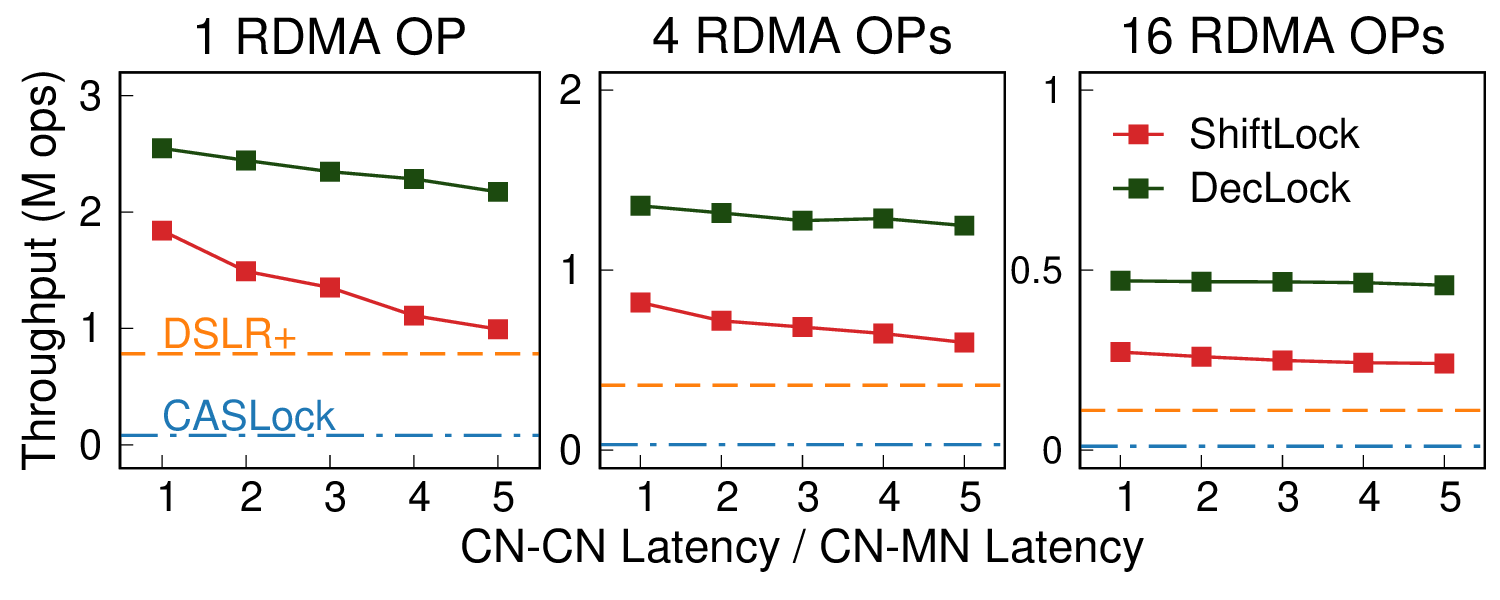}
  \end{minipage} \\[-8pt]
  \begin{minipage}{1.\linewidth}
    \caption{\small{\emph{The throughput of the microbenchmark in a
    heterogeneous network with different CN-CN communication latencies 
    when using different lock mechanisms.
    }}}
    \label{fig:hetero}
  \end{minipage}  \\[-0pt]
\end{figure}

To simulate a heterogeneous network where CN-CN communication has higher
latency than CN-MN communication, we manually insert delays to
RDMA operations between CNs and run the microbenchmark with varying ratios 
of CN-CN to CN-MN latency. We use the default workload parameters described
in \S\ref{sec:eval:micro}, except for adjusting the critical section length to 1, 4, and 16
RDMA operations. {\sys} uses the phase-fair design across all experiments.

As shown in Fig.~\ref{fig:hetero}, the throughput of CASLock and DSLR+ is not affected
by CN-CN latency because all clients acquire locks by simply checking
or updating lock states on MNs. The throughput of {\sys} and ShiftLock
reduces by at most 14.7\% and 49.6\%, respectively, due to their reliance
on CN-CN notifications. ShiftLock exhibits a more significant throughput
degradation as CN-CN latency increases because its protocol requires
twice as many CN-CN messages as {\sys}. 

As the critical section length
increases, the impact of CN-CN latency on {\sys}'s throughput diminishes
because the ownership transfer between CNs becomes less frequent.
Even with the shortest critical sections (1 RDMA operation), {\sys} still
outperforms DSLR+ and CASLock by at least 2.79\x and 27.18\x, respectively.

\end{appendix}